\documentclass[12pt]{article}

\usepackage{pifont}
\newcommand{\cmark}{\ding{51}}%
\newcommand{\xmark}{\ding{55}}%

\usepackage{latexsym}
\usepackage{amsthm}
\usepackage{amsmath,amssymb,amsfonts,bm,amsbsy,bbm}
\usepackage{epsfig}
\usepackage{graphicx,rotating,lscape}
\usepackage{verbatim}
\usepackage{multirow,color}
\usepackage{geometry}
\usepackage{setspace}
\usepackage{subfigure}
\usepackage{bbm}
\usepackage{ulem} 
\newcommand{\stkout}[1]{\ifmmode\text{\sout{\ensuremath{#1}}}\else\sout{#1}\fi}

\usepackage{enumerate}
\usepackage{url} 
\usepackage{tikz, adjustbox, float} 
\usepackage{dsfont} 
\usepackage{mathrsfs} 
\usepackage{soul} 
\usepackage{bm} 
\usepackage{booktabs}  
\usepackage{enumitem} 

\usepackage[ruled,vlined]{algorithm2e}
\SetKwInOut{Input}{Input}
\SetKwInOut{Output}{Output}
\SetKwRepeat{Do}{do}{while}%

\usepackage[linktoc=all,colorlinks=true,linkcolor=black,citecolor=blue,urlcolor=blue,bookmarks=false,hypertexnames=true]{hyperref}

\def\real{{\rm I\!R}}

\def\0{{\bf 0}}
\def\1{{\bf 1}}

\def\btheta{{\bm{\theta}}}
\def\htheta{\widehat{\theta}}
\def\hbtheta{\widehat{\bm{\theta}}}

\def\bSigma{\bm{\Sigma}}
\def\hsigma{\widehat{\sigma}}

\def\blambda{{\bm{\lambda}}}
\def\bLambda{{\bm{\Lambda}}}

\def\I{{\bf I}}

\DeclareMathOperator*{\var}{var}  
 
\DeclareMathOperator*{\argmax}{argmax}
\DeclareMathOperator*{\argzero}{argzero}
\DeclareMathOperator*{\argsup}{argsup}

\DeclareMathOperator*{\diag}{diag}

\def\bt{{\boldsymbol\theta}}

\def\z{{\bf z}}

\def\e{{\bf e}}
\def\c{{\bf c}}
\def\t{{\bf t}}

\def\0{{\bf 0}}
\def\trans{^{\rm T}}

\def\wh{\widehat}
\def\wt{\widetilde}

\def\n{\nonumber}

\def\var{\hbox{var}}

\def\argmax{\mbox{argmax}}
\def\diag{\mbox{diag}}

\def\bse{\begin{eqnarray*}}
\def\ese{\end{eqnarray*}}
\def\be{\begin{eqnarray}}
\def\ee{\end{eqnarray}}
\def\bsq{\begin{equation*}}
\def\esq{\end{equation*}}
\def\bq{\begin{equation}}
\def\eq{\end{equation}}

\def\boxit#1{\vbox{\hrule\hbox{\vrule\kern6pt  \vbox{\kern6pt#1\kern6pt}\kern6pt\vrule}\hrule}}
\def\wh{\widehat}
\def\wt{\widetilde}

\def\argmax{\mbox{argmax}}
\def\diag{\mbox{diag}}
\def\sumi{\sum_{i=1}^n}

\def\trans{^{\rm T}}

\def\calA{{\cal A}}

\newtheoremstyle{mytheoremstyle} 
    {0.3cm}                      
    {0cm}                        
    {}
    {}                           
    {\bf}                   
    {: }                          
    {0em}                       
    {}  

\newtheoremstyle{myExampleRemarkstyle} 
    {0.3cm}                    
    {0cm}                      
    {\it}                         
    {}                         
    {\bf}                      
    {: }                       
    {0em}                      
    {}  

\theoremstyle{myExampleRemarkstyle}

\newtheorem{pro}{\bf Proposition}
\newtheorem{cor}{Corollary}
\newtheoremstyle{cor} 
    {0.3cm}                      
    {0cm}                        
    {}
    {}                           
    {\bf}                   
    {: }                          
    {0em}                       
    {}  
\theoremstyle{definition}

\theoremstyle{remark}

\newtheoremstyle{remark} 
    {0.3cm}                    
    {0.3cm}                      
    {\it}                         
    {}                         
    {\bf}                      
    {: }                       
    {0em}                      
    {}  
\newtheorem{theorem}{Theorem}
\newtheoremstyle{example}{}{}{}{}{\bfseries}{.}{ }{}
\theoremstyle{example}

\newtheorem{lemma}{Lemma}

\newtheorem{innercustomgeneric}{\customgenericname}
\providecommand{\customgenericname}{}
\newcommand{\newcustomtheorem}[2]{%
  \newenvironment{#1}[1]
  {%
   \renewcommand\customgenericname{#2}%
   \renewcommand\theinnercustomgeneric{##1}%
   \innercustomgeneric
  }
  {\endinnercustomgeneric}
}

\newcustomtheorem{customthm}{Theorem}
\newcustomtheorem{customlemma}{Lemma}
\usepackage{mathalfa}
\usepackage{oubraces}

\let\refBKP\ref
\renewcommand{\ref}[1]{{\upshape\refBKP{#1}}}

\newcommand{\blind}{0}

\usepackage{natbib}
\setcitestyle{citesep={,}}
\setcitestyle{aysep={}} 

\begin{document}

\def\spacingset#1{\renewcommand{\baselinestretch}%
{#1}\small\normalsize} \spacingset{1}

\if1\blind
{
  \begin{center}
    {\LARGE\bf Bioequivalence Assessment for Locally Acting Drugs: A Framework for Feasible and Efficient Evaluation}
\end{center}
} \fi

\if0\blind
{
  \begin{center}
    {\LARGE\bf Bioequivalence Assessment for Locally Acting Drugs: A Framework for Feasible and Efficient Evaluation}

\vskip 0.5cm
$\textsc{Luca Insolia}^{\text{1-3}}$, 
$\textsc{Yanyuan Ma}^{\text{4}}$, \\
$\textsc{Younes Boulaguiem}^{\text{5}}$ 
\& \textsc{St\'ephane Guerrier}$^{\text{1-3},\star}$ 
\end{center}

\begin{center}
 \footnotesize
    $^{1}$School of Pharmaceutical Sciences, University of Geneva, Geneva, Switzerland;
    $^{2}$Institute of Pharmaceutical Sciences of Western Switzerland, University of Geneva, Geneva, Switzerland;
    $^{3}$Department of Earth Sciences, University of Geneva, Geneva, Switzerland;
    $^{4}$Department of Statistics, The Pennsylvania State University, University Park, Pennsylvania, USA;
    $^{5}$Geneva School of Economics and Management, University of Geneva, Geneva, Switzerland.
\end{center}  
  \medskip
} \fi

\bigskip
\begin{abstract}
\noindent
Equivalence testing plays a key role in several domains, such as the development of generic medical products, which are therapeutically equivalent to brand-name drugs but with reduced cost and increased accessibility.
Promoting access to generics is a critical public health issue with substantial societal implications, but establishing equivalence is particularly challenging in multivariate settings. A notable example refers to locally acting drugs designed to exert their therapeutic effects at a localized area where they are administered rather than being absorbed into the bloodstream, where complex experimental protocols lead to reduced sample sizes and substantial experimental noise. Traditional approaches, such as the Two One-Sided Tests (TOST), cannot adequately tackle the complex multivariate nature of such data. 
In this work, we develop an adjustment for the TOST procedure by simultaneously correcting its significance level and equivalence margins to ensure control of the test size and increase its power. In large samples, this approach leads to an optimal adjustment for the univariate TOST procedure. In multivariate settings, where we show that an optimal adjustment does not exist, our proposal maintains equal marginal test sizes and overall size control while maximizing power in important cases.
Through extensive simulation studies and a case study on multivariate bioequivalence assessment for two antifungal topical products, we demonstrate the superior performance of our method across various scenarios encountered in practice. 
\end{abstract}

\noindent
{\it Keywords:} 
composite hypothesis testing, equivalence test, finite-sample adjustments, locally acting drugs, two one-sided tests.

\spacingset{1.9} 

\section{Introduction}

Average equivalence testing aims at detecting whether an average effect of
interest, such as the difference in two means, falls within a predetermined region of practical equivalence. 
Nowadays, there is an ever-increasing interest in equivalence testing problems, especially in multivariate contexts \citep{pallmann2017simultaneous}. These problems arise in many
disciplines, such as psychology \citep{Lakens:18}, engineering
\citep{moore2022engineering}, software development
\citep{dolado2014equivalence}, and social sciences
\citep{aggarwal20232}, to name a few.
Equivalence testing plays a prominent role in the pharmaceutical sciences, where it is typically referred to as bioequivalence testing.
Bioequivalence studies play a key role in developing generic
drugs, which are therapeutically equivalent to brand-name drugs
but with reduced cost and increased medication accessibility.
Promoting the accessibility of generic medical products is an important public health concern with significant social and economic effects
(see e.g.,~\citealp{lukic2020comparison}, \citealp{shin2020cutaneous}, \citealp{miranda2022topical}).  A key aspect of the approval process
for generics involves demonstrating their therapeutic equivalence to
well-investigated brand-name drugs
(see e.g.,~\citealp{patterson2017bioequivalence}). As most
administered drugs enter the bloodstream to reach their site of
action, their assessment traditionally relies on comparing 
the active
ingredient concentrations in the blood using
pharmacokinetic parameters (e.g.,~the area under the blood
concentration-time curve or the peak concentration). A test drug (such
as a new generic drug)
and a reference drug (such as a brand-name drug)
are considered \textit{bioequivalent}
if their rate and extent of absorption, as quantified by pharmacokinetic
parameters, are sufficiently close \citep{du2015likelihood}. Several
approaches have been considered for the assessment of bioequivalence (see 
e.g.,~\citealt{MEYNERS2012} for a review), 
among which the Two One-Sided Tests (TOST) procedure has emerged as
the most widely employed approach
\citep{berger1996bioequivalence}. This experimental and statistical
framework is recognized by the United States Food and Drug
Administration (FDA), the European Medicines Agency (EMA), and other
regulatory authorities for the approval of generic drugs or new formulations of existing drugs.
Importantly, bioequivalence
is also used in other critical situations throughout the drug development life-cycle (e.g.,~pre-clinical and clinical development, manufacturing and quality control, pharmacovigilance, etc.).

In this work, we propose a new approach for multivariate equivalence testing, an area where few methods have been developed. A key motivation for this new approach is the assessment of bioequivalence for locally acting drugs, which are medications
designed to exert their therapeutic effects at the site of application
without significant systemic absorption.  These drugs are not intended
to be absorbed into the bloodstream and have local sites of action
such as the skin, the eyes, the ears, or the mucosal surfaces of the
nose and the lungs, minimizing systemic side effects and drug interactions and thus contributing to safer and more effective therapies. The assessment of bioequivalence for locally acting
drugs is crucial for researchers, regulatory agencies, and generic
drug manufacturers  (see e.g.,~\citealp{patil2021overview}, \citealp{mehta2023global},
\citealp{babiskin2023regulatory}). However, establishing
bioequivalence for these drugs poses significant challenges,
since
pharmacokinetic parameters, which are crucial for systemically acting
drugs, are only indicative of unintended side effects and are not
reliable metrics for assessing bioequivalence in this context (see
e.g.,~\citealp{herkenne2008vivo}, \citealp{cordery2017topical}), leading to more complex, labor-intensive, and time-consuming experimental protocols and thus to smaller sample sizes.
Consequently,
there are relatively fewer generic products for locally acting drug
products, and there is an urgent need for adequate experimental and
statistical methods to address this challenge
\citep{miranda2018bioequivalence}. Recently, this limitation was
recognized by the FDA and the EMA, which have called for innovative
methods to evaluate bioequivalence for locally acting drugs,
particularly topical formulations
(see \citealp{EMA2018}, \citealp{FDA2022}, \citealp{volonte2024equivalence}). 
Indeed, the complexity of the underlying clinical protocols leads to large experimental noise which, 
combined with the multivariate nature of such data,
renders classical bioequivalence methods, such as the TOST, to have
virtually nil statistical power. 

\subsection{Main results and contributions}

In this paper, we propose a novel finite-sample adjustment for multivariate average equivalence testing problems, which also applies to univariate settings as a special case.
To mitigate the conservativeness of the TOST procedure while preserving its simplicity, our proposal matches its test size to the nominal significance level by simultaneously adjusting both the test level and the equivalence margins in use. The proposed approach, which we denote as \textit{corrected TOST} (cTOST), leads to a test procedure with superior operating characteristics compared to the conventional TOST and other existing methods.

In univariate settings, we show that all adjusted TOST procedures are equivalent when the nuisance parameter (i.e.,~the variance) is assumed to be known. 
When this parameter is unknown but still employed to characterize the theoretical test size of TOST, we show that the proposed cTOST leads to the optimal adjustment, in the sense that it leads to a uniformly most powerful test procedure. 
In practice, when the nuisance parameter is replaced by a plugin estimate, we demonstrate that the resulting cTOST maintains such optimal properties in large samples.
Nevertheless, in the presence of very small sample sizes, we observe that it may lead to a slightly liberal test procedure. We thus propose a further small-sample refinement that ensures control of the test size while still maintaining desirable properties.

Building on our findings from the univariate case, we extend our focus to the more complex multivariate setting. We demonstrate that no uniformly most powerful test exists and that locally optimal solutions often yield impractical rejection regions.
We thus propose a constrained optimization framework that balances theoretical optimality with practical considerations. By controlling the marginal test sizes across dimensions, our approach yields a computationally efficient adjustment that is both optimal in important cases and practically meaningful.
Moreover, the proposed small-sample refinement can be also applied to the multivariate setting.

Finally, we apply the proposed cTOST to multivariate bioequivalence assessments in the context of locally acting drugs, where the underlying labor-intensive and time-consuming experimental protocols severely limit the amount of available data.
Unlike existing methods, cTOST can effectively tackle the large experimental noise and achieve a declaration of bioequivalence when other methods fail. 
We present a real data example of the cutaneous biodistribution between two antifungal topical products for different targeted skin layers \citep{quartier2019cutaneous}. 
This case study illustrates the lack of statistical power in classical bioequivalence testing procedures and showcases the superior operating characteristics achieved by cTOST. 

\if0\blind{
All our new testing procedures are
readily available for public usage through the \texttt{cTOST} package
in \texttt{R}, which is also available on the GitHub repository
\url{https://github.com/stephaneguerrier/cTOST}. 
}\fi
\if1\blind{
All our new testing procedures are
readily available for public use through GitHub. 
}\fi

\subsection{Organization}

The rest of the article is organized as
follows. Section~\ref{sec:univariate} briefly reviews the framework for multivariate equivalence testing as well as the TOST and
other existing procedures, considering the univariate setting as a special case. 
In Section~\ref{sec:proposed}, we introduce our cTOST proposal for univariate and multivariate equivalence testing problems, where we also
demonstrate its theoretical properties and connections with existing methods, and present a further small-sample refinement.
Section~\ref{sec:simulation} includes
extensive simulation studies comparing the performance of different
methods, and Section~\ref{sec:application} contains the assessment of bioequivalence for two antifungal topical products. 
Finally, some concluding remarks are discussed in
Section~\ref{sec:conclusion}. 

\section{Average Equivalence Testing Framework}
\label{sec:univariate}

To demonstrate average equivalence, the hypotheses of interest are given by
\be \label{eqn:equiv_hyp_glob_mvt}
    \text{H}_0: \; \bt \not\in ( -c_0, \,
    c_0)^K \quad \text{vs.} \quad
    \text{H}_1: \; \bt \in ( -c_0, \,
    c_0)^K,
\ee
where $\bt$ denotes the target equivalence $K$-dimensional vector and 
$c_0$ is a given constant based
on expert domain knowledge. For example, in bioequivalence studies, this value could be a
standard set by regulatory agencies allowing to quantify how ``close'' two drugs must be to be declared bioequivalent. Without loss of generality, 
we restrict our attention to the conventional choice of equivalence
margins that are equal across the $K$ components and symmetric around zero (see
e.g.,~\citealp{berger1996bioequivalence}, \citealp{pallmann2017simultaneous}
and the references therein). 

To test the hypotheses in \eqref{eqn:equiv_hyp_glob_mvt}, a common assumption (see
e.g.,~\citealp{schuirmann1987comparison}, \citealp{jones2003design}, \citealp{wang1999statistical}) 
is that an
estimator of $\btheta$, say $\hbtheta$, is available and satisfies the
following canonical form:
\be\label{eq:mvt_canon}
    \wh\bt\sim \mathcal{N}_K(\bt, \bSigma_1)  \;\;\; \text{and}\;\;\;  \nu_2
    \wh\bSigma_1 \sim W_K(\bSigma_1,\nu_2) ,
\ee
where the two random variables respectively following the normal and
the Wishart distribution are independent. Here
$\bSigma_1\equiv\bSigma/\nu_1$ denotes the covariance matrix of
$\wh\bt$, which we assume to be positive definite, and $\nu_1>0, \nu_2>0$ are degrees of freedom related to the
sample size, $\nu_1\asymp\nu_2\asymp n$, 
 where $u_n \asymp v_n$ indicates that there exist positive
constants $m_l$, $m_u$ and $N$ such that 
$m_l \leq \lvert u_n / v_n \rvert \leq m_u $ for all $n > N$.
For example,
when $K=1$ and the data $X_1, \dots, X_n$ satisfy 
$E(X_i)=\theta$, then we usually have 
$\wh\theta=\overline X = n^{-1} \sum_{i=1}^n X_i \sim \mathcal{N} (\theta, \sigma_1^2) $,
$\sigma^2 \equiv \Sigma =\var(X)$, $\sigma_1^2=\var(X)/n$,
$\nu_1=n$, $\nu_2=n-1$,
$\wh\sigma_1^2=\sumi(X_i-\overline X)^2/\{n(n-1)\}$, and 
$\nu_2 \wh\sigma_1^2 / \sigma_1^2 \sim \chi^2_{\nu_2}$ (where the latter denotes a chi-square distribution with $\nu_2$ degrees of freedom).
More complex (regression) settings can also be summarized into the above
form (see e.g.,~\citealp{counsell2015equivalence} and the references
therein) either exactly or in an asymptotic sense.

Several approaches have been considered for univariate average equivalence testing (i.e.,~when $K=1$; see
e.g.,~\citealt{MEYNERS2012} for a review), 
among which TOST procedure has emerged as
the most widely adopted approach
(\citealp{schuirmann1987comparison}, \citealp{berger1996bioequivalence}). This
procedure requires to compute the following two test statistics:  
$
T_L \equiv (\widehat{\theta}+c_0) / \wh\sigma_1
$
and 
$
T_U \equiv (\widehat{\theta}-c_0) / \wh\sigma_1,
$
where $T_L$ tests for $\text{H}_{01}: \theta\leq -c_0$ versus
$\text{H}_{11}: \theta> -c_0$, and $T_U$ tests for $\text{H}_{02}:
\theta \geq c_0$ versus $\text{H}_{12}: \theta<c_0$. At a significance
level $\alpha_0 \in (0, 1/2]$, the TOST rejects
$\text{H}_{0}\equiv\text{H}_{01} \cup \, \text{H}_{02}$  in favor of
$\text{H}_{1}\equiv\text{H}_{11} \cap \, \text{H}_{12}$ if both tests
simultaneously reject their marginal null hypotheses, that is, 
if $T_L \geq t_{\alpha_0,\nu_2}$ and $T_U \leq -t_{\alpha_0,\nu_2}$,
where $t_{\alpha_0,\nu_2}$ denotes the upper $\alpha_0$ quantile of a
$t$-distribution with $\nu_2$ degrees of freedom. 
Both tests are designed to control the false positive rate (type I
error rate) at $\alpha_0$.
 Hence, the
intersection-union principle guarantees that the resulting TOST
procedure is indeed level $\alpha_0$
\citep{berger1996bioequivalence}.
In practice, the most common way of assessing (and visualizing)
bioequivalence is to use the Interval Inclusion Principle (IIP) and
accept bioequivalence if the $1-2\alpha_0$ confidence interval for 
$\theta$ falls entirely
within the equivalence margins $(-c_0, c_0)$;  see
e.g.,~\citet{wellek2010testing}. 

The rejection region of the (univariate) TOST procedure is given by
\bse
    C \equiv \left\{\wh\theta \in \real,
      \wh\sigma_1 \in \real_{>0} :\,
t_{\alpha_0,\nu_2} \wh\sigma_1-c_0<
\wh\theta<c_0-t_{\alpha_0,\nu_2} \wh\sigma_1\right\}.
\ese
Therefore, given $\alpha_0$, $\theta$, $\sigma_1$, $\nu_2$ and $c_0$,
the probability of rejecting $\text{H}_0$ can be expressed as
\citep[see e.g.,][]{phillips2009power}
\be
\begin{aligned}\label{eq:tost_power_fn}
    & \omega(\theta, \sigma_1, \nu_2, t_{\alpha_0,\nu_2}, c_0) \equiv\Pr(t_{\alpha_0,\nu_2} \wh\sigma_1-c_0<\wh\theta<c_0-t_{\alpha_0,\nu_2} \wh\sigma_1)\\
    &=\int I(\wh\sigma_1t_{\alpha_0,\nu_2} < c_0)\left\{\Phi\left(\frac{c_0-t_{\alpha_0,\nu_2} \wh\sigma_1-\theta}{\sigma_1}\right)
-\Phi\left(\frac{t_{\alpha_0,\nu_2} \wh\sigma_1-c_0-\theta}{\sigma_1}
\right) \right\}f_{\wh \sigma_1}( \wh\sigma_1 \lvert \sigma_1,\nu_2)
        d\wh\sigma_1 ,
\end{aligned}
\ee
where $I(\cdot)$ denotes the indicator function with $I(A) = 1$ if $A$ is true and zero otherwise, 
$\Phi(\cdot)$ denotes the cumulative distribution function of a
standard normal distribution, and $f_{\wh \sigma_1}(\cdot) $ is the
density of $\wh \sigma_1$.
Subsequently,
the size of the test is defined as the supremum of
  $\omega(\theta, \sigma_1, \nu_2, t_{\alpha_0,\nu_2}, c_0)$ over the
space of the 
null hypothesis (see e.g.,~\citealp{Lehmann:1986}). It can be shown that for a
given $\sigma_1$ we have 
\be \label{eq:tost_size}
\sup_{\theta\not\in  (-c_0, c_0)} \,\omega(\theta,\sigma_1, \nu_2,
    t_{\alpha_0,\nu_2}, c_0) 
    = \sup_{\theta \in [c_0, \infty)}\,\omega(\theta,\sigma_1, \nu_2,
    t_{\alpha_0,\nu_2}, c_0) 
    = \omega(c_0, \sigma_1, \nu_2, t_{\alpha_0,\nu_2}, c_0) < \alpha_0,
\ee
indicating that  TOST is a conservative procedure in the sense that
its size is strictly smaller than the target level $\alpha_0$
(see e.g.,~\citealp{Deng2020}). A few strategies have been considered
to tackle the conservativeness
of the univariate TOST procedure. In the context of bioequivalence, some approaches considered the
adjustments of regulatory margins;
see for instance
\citet{tothfalusi2009evaluation,davit2012implementation,munoz2016consumer,labes2016inflation,schutz2022critical}. 
However, these
approaches tend to provide overly liberal test procedures while maintaining a limited power \citep{ocana2019controlling}.
Recently, the $\alpha$-TOST and  $\delta$-TOST procedures were
proposed in \cite{boulaguiem2024finite} to 
consider either an adjusted test level, say $\alpha^*$, or adjusted
equivalence margins, say $c^*$, respectively.
However, these adjustments only partially mitigate the conservativeness of TOST and remain suboptimal.
Alternatives to  TOST could also be considered, but they tend to give
overly liberal testing procedures, can be difficult to compute 
\citep{anderson1983new}, and are typically difficult to interpret as
they do not respect the IIP \citep{berger1996bioequivalence}. 

Notably, fewer approaches have been proposed for assessing average
equivalence in multivariate settings  
(i.e.,~when $K>1$; see e.g.,~\citealp{tseng2002optimal},
\citealp{chervoneva2007multivariate}, \citealp{wellek2010testing} 
and the references therein). Among these,
the multivariate TOST procedure stands
out as the most effective method as 
demonstrated by 
\citet{pallmann2017simultaneous} 
empirically. 
It relies on test statistics:  
$
T_{L_k} \equiv (\htheta_k + c_0) / \wh\sigma_{1,k}
$
and
$
T_{U_k} \equiv (\htheta_k - c_0) / \wh\sigma_{1,k} ,
$
for $k=1,\ldots,K$, where $\wh\sigma_{1,k}^2$ is the $k$th diagonal
element of $\wh\bSigma_1$,  and
 rejects H$_0$ if each of the $K$ marginal TOSTs rejects its marginal
null hypothesis.
Similarly to the univariate case, the IIP ensures that equivalence can be declared when each marginal
confidence interval for $\theta_k$, where $k=1, \ldots, K$, is entirely contained within the equivalence margins $(-c_0, c_0)$.
At a significance level $\alpha_0$, 
the multivariate TOST has the rejection region of the form 
\bse
C_K \equiv \left\{ \wh\bt \in \real^K,
      ( \wh\sigma_{1,1}, \ldots, \wh\sigma_{1,K})\trans  \in \real_{>0}^K :\,
\bigcap_{k=1}^K \left( t_{\alpha_0,\nu_2} \wh\sigma_{1,k}-c_0<\wh\theta_k< c_0 - t_{\alpha_0,\nu_2} \wh\sigma_{1,k} \right)\right\} ,
\ese
which is derived from an extension of the univariate TOST to the
multivariate framework \citep{berger1982interunion}.
Then, the probability of rejecting H$_0$ for
the multivariate TOST (see e.g.,~\citealp{phillips2009power}) can be expressed as 
$
 \omega_K (\bt, \bSigma_1, \nu_2 , t_{\alpha_0, \nu_2},  c_0 ) \equiv
  \Pr ( \wh\btheta \in C_K 
    \lvert 
    \btheta, \bSigma_1, \nu_2 , \alpha_0, c_0
    )$.
While the multivariate TOST
remains easier to interpret and exhibits greater power compared to
other testing procedures \citep{pallmann2017simultaneous}, its conservativeness persists and in fact becomes more pronounced when applied to 
multivariate equivalence 
problems. This increased conservativeness suggests improvement
opportunities and motivates the development of further refinements.

\section{Towards an Optimally Corrected TOST: cTOST}
\label{sec:proposed}

In this section, 
we detail the derivation and theoretical properties of the proposed cTOST, its advantages over existing methods and its practical implications. 
While the aim is optimality, the multivariate context presents unique challenges.
Therefore, we first focus on the univariate setting, demonstrating that the proposed cTOST leads to an optimal adjustment.  
We then extend our approach to the more complex multivariate framework where, although we establish that an optimal adjustment does not exist, cTOST maintains a degree of optimality in important cases. 
Moreover, to control its properties in small samples, we introduce an additional refinement step ensuring that cTOST does not lead to a liberal test procedure.

\subsection{Univariate cTOST}
\label{sec:proposed:univariate}

We aim at constructing an optimal adjustment for the TOST, in the
sense that it achieves a size equal to the nominal significance level
$\alpha_0$ while simultaneously maximizing the power of the test
-- i.e.,~the probability of rejecting H$_0$ when $ \theta \in (-c_0, c_0)$.
To achieve this, we consider a more flexible rejection region by
treating $c_0$ and the $\alpha_0$ in $t_{\alpha_0,\nu_2}$ as
parameters $c,\alpha$ to get
\be\label{eq:region}
C(\alpha,c)\equiv \left\{\wh\theta \in \real,
      \wh\sigma_1 \in \real_{>0} :\,
t_{\alpha,\nu_2} \wh\sigma_1-c<
\wh\theta<c-t_{\alpha,\nu_2} \wh\sigma_1\right\},
\ee
and then match the size function in \eqref{eq:tost_size} to
the nominal level $\alpha_0$ by requiring that $\omega (c_0, \sigma_1, \nu_2, t_{\alpha,\nu_2}, c)=\alpha_0$. This
results in a set $\calA(\alpha_0)$ of possible  adjustments, formally
defined as 
\be \label{eq:adjustments}
\calA(\alpha_0) \equiv\{(\alpha, c):
\alpha \in (0,0.5], c \in \real_{>0},  \omega(c_0, \sigma_1, \nu_2,
t_{\alpha,\nu_2}, c )=
\alpha_0 \}.
\ee
We note that the size function $\omega (c_0, \sigma_1, \nu_2, t_{\alpha,\nu_2}, c)$ has the property that, at any fixed $t_{\alpha,\nu_2}\in[0,\infty)$, 
it is a strictly increasing and differentiable function of $c$, and
has value 0 at $c=0$ and value 1 as $c\to\infty$. Hence, for any
$t_{\alpha,\nu_2}\in[0,\infty)$, there exists a unique value of $c$,
denoted as $c(t_{\alpha,\nu_2},\nu_2)$, so that $\omega\{c_0,
\sigma_1, \nu_2, t_{\alpha,\nu_2}, c(t_{\alpha,\nu_2},\nu_2)\}
=\alpha_0$.
Thus, $\calA(\alpha_0)=
  \{(\alpha, c(t_{\alpha,\nu_2},\nu_2)\}$, where $\alpha \in (0,0.5]$,
  and we aim at finding the optimal $\alpha$ so that the power $\omega
  \{\theta, \sigma_1, \nu_2, t_{\alpha,\nu_2}, c(t_{\alpha,\nu_2},\nu_2)\}$
  is maximized for 
  $\theta \in (-c_0, c_0)$.
  Finding such optimal $\alpha$ is
  equivalent to finding the corresponding optimal $t_{\alpha, \nu_2}$ value,
  so we simplify the above problem as finding $t$ so that
  $\omega
  \{\theta, \sigma_1, \nu_2, t, c(t,\nu_2)\}$
  is maximized for 
   $\theta \in (-c_0, c_0)$.
  The solution is surprisingly simple,
  in that we find the optimal $t$ is simply zero for all
   $\theta \in (-c_0, c_0)$.
  Specifically, for notational simplicity,
  we define
\be\label{eq:xtost_matching}
    c(0) \equiv c(0, \nu_2)= \argzero_{ c \in \real_{>0} }\ \{ \omega(c_0, \sigma_1, \nu_2, 0, c)-\alpha_0 \} .
\ee
We find that setting $c$ at $c(0)$  leads to the optimal power.  
While the result is formally given in Theorem
\ref{thm:most_powerful}, we now provide some intuition behind.
In the $\alpha$-TOST procedure,  which also belongs to the set of adjustments $\calA(\alpha_0)$ in \eqref{eq:adjustments}, 
it was shown that an increase in
$\alpha$
leads to more powerful test procedures compared to an
increase in  $c$.
The maximum value at $\alpha=1/2$
is then a natural
candidate.  
Moreover, the power function at $\alpha=1/2$, which is $\omega(\theta,
\sigma_1, \nu_2, 0, c)$, 
coincides with the power function for the known $\sigma_1$ case (see
Appendix~\ref{appendix:power_known_sigma}), which intuitively is the
upper bound of the power of the unknown $\sigma_1$ case. 
On the other hand, when $\sigma^2_1$ is known, all
adjustments that lead to size-$\alpha_0$ procedures have equal
power under H$_1$ (see Appendix~\ref{appendix:power_known_sigma} for
details),  and they are equivalent to the uniformly most powerful test
presented in  \citet{romano2005optimal}.
This observation indicates that the adjustment
in \eqref{eq:xtost_matching} leads to the uniformly most powerful test
procedure, as demonstrated in Theorem~\ref{thm:most_powerful}.
\begin{theorem} \label{thm:most_powerful}
   For any given $\nu_2$ and $\sigma_1$,
    $c(0)$
    defined in \eqref{eq:xtost_matching} satisfies
    $
    \omega \{ \theta, \sigma_1, \nu_2, t, c(t,\nu_2) \}
    \leq 
    \omega \{ \theta, \sigma_1, \nu_2, 0, c(0) \} ,
    $
   for any $t > 0$,
   where $\{ t, c(t, \nu_2)\} \in \calA(\alpha_0) $ in \eqref{eq:adjustments},
   and any $\theta \in (-c_0, c_0)$. 
\end{theorem}
 
The proof of Theorem~\ref{thm:most_powerful} is presented in
Appendix~\ref{appendix:_power_xtost_univ}.
Theorem~\ref{thm:most_powerful} indicates that among all
level-$\alpha_0$ tests, cTOST is the uniformly most powerful test. 
In addition to its superiority in terms of power, cTOST offers
substantial computational gain compared to the existing adjusted TOST
procedures, which often require the computation of an additional
integral over the space of $\hsigma_1$
as given in \eqref{eq:tost_power_fn}. 
Although Theorem~\ref{thm:most_powerful} considers the case that
$\sigma_1$ is unknown, the resulting solution, i.e.,~the optimal
choice $c(0)$, depends on the true $\sigma_1$ value, 
which in practice we replace by
\be\label{eq:whc0}
\wh c(0)\equiv \argzero_{ c \in \real_{>0} }\ \{ \omega(c_0,
\wh\sigma_1, \nu_2, 0, c)-\alpha_0\} .
\ee
For simplicity, we still refer to the resulting
procedure as cTOST even though it now relies on $\hsigma_1$ instead of $\sigma_1$.
Similarly, we let $\wh c(t,\nu_2)$ 
satisfy $\omega\{c_0,
\wh \sigma_1, \nu_2, t_{\alpha,\nu_2}, \wh c(t,\nu_2)\} = \alpha_0$.
To show that the substitution of $\sigma_1$ by $\wh\sigma_1$
does not harm the performance of cTOST and that it remains the optimal adjustment in large
samples, we first establish some preliminary results in
Proposition~\ref{pro:asymp}. 
\begin{pro} \label{pro:asymp}
For any $t>0$, we have:
\begin{enumerate}
    \item $\wh c(0)-c(0)=O_p(\sigma_1/\sqrt{\nu_2}),$
    \item $\wh c(t, \nu_2)-c(t, \nu_2)=O_p(\sigma_1/\sqrt{\nu_2}),$
    \item $\lvert c(0)-\{c(t, \nu_2)-t\wh \sigma_1\} \rvert\asymp \sigma_1 .$
\end{enumerate}
\end{pro}
The proof of Proposition~\ref{pro:asymp} is given in
Appendix~\ref{appendix:pro_asymp}. 
Proposition~\ref{pro:asymp} 
indicates that for large $n$,
the difference between $\wh c(0)$ and $c(0)$ (result 1)
caused by estimating $\sigma_1$,
as well as the difference between $\wh c(t,\nu_2)$ and $c(t,\nu_2)$
for a general $t$ (result 2),
are ignorable compared to the
difference caused by choosing a nonzero $t$ versus a zero $t$ value
(result 3, see \eqref{eq:region}),
which is what cTOST does.
Thus, the $\wh c(0)$ adjustment still has the optimal power
property while retaining the size $\alpha_0$ in large samples. This result is formally shown in the following theorem.
\begin{theorem}\label{th:asy}
$\wh c(0)$ defined in \eqref{eq:whc0} satisfies:
\begin{enumerate}
\item
For any $t \geq 0$,
  $\omega \{ c_0, \sigma_1, \nu_2,
      t, \wh c(t,\nu_2) \} = \alpha_0 + O_p(\nu_2^{-1/2})$.
    \item
For any $t > 0$, 
        and any $\theta \in (-c_0, c_0)$, when $\nu_2$ is
        sufficiently large,
      $ \omega \{ \theta, \sigma_1, \nu_2, t, \wh c(t,\nu_2) \}
    \leq 
    \omega \{ \theta, \sigma_1, \nu_2, 0, \wh c(0) \}
    $.
\end{enumerate}
\end{theorem}
The proof of Theorem~\ref{th:asy} is given in 
Appendix~\ref{appendix:th:asy}. 
The first part of Theorem~\ref{th:asy} implies that the asymptotic
size of the corrected procedure is $\alpha_0$.  
The second part of this result shows that  with only an estimated
$\sigma_1$ available, for large $n$,
setting $\alpha$ to $1/2$
still leads to the optimal adjustment.
In practice, $\wh c(0)$ can be efficiently computed using, for
instance, the Newton-Raphson method described in
Appendix~\ref{appendix:algo_ctost}.
Then, if $|\wh 
\theta| < \wh c(0)$, we accept H$_1$ at the significance level
$\alpha_0$.

\subsection{Multivariate cTOST}
\label{sec:proposal_multiv}

Similarly to the univariate case, it is easy to see that the
multivariate TOST is conservative, meaning that its size, which can
be derived from \eqref{eq:adjustments_mvt}-\eqref{eq:lambda_argsup} below, is strictly smaller than the nominal level $\alpha_0$.
Following the univariate cTOST idea, for each $k=1, \dots, K$,
we relax the marginal test rejection regions to $\{\wh \theta_k, \wh\sigma_{1,k}:
t\wh\sigma_{1,k}-c<\wh\theta<c-t\wh\sigma_{1,k}\}$. This leads to the
multivariate test rejection region of the form
\bse
C_K (\t, \c) \equiv \left\{ \wh\bt \in \real^K,
      ( \wh\sigma_{1,1}, \ldots, \wh\sigma_{1,K})\trans  \in \real_{>0}^K :\,
\bigcap_{k=1}^K \left( 
t_k\wh\sigma_{1,k}-c_k<\wh\theta_k< c_k-t_k\wh\sigma_{1,k}
\right)\right\},
\ese
where
$\t\equiv(t_1, \dots, t_K)\trans$ 
and $\c\equiv(c_1, \dots, c_K)\trans$.
We denote the probability of rejecting H$_0$ as
\be \label{eq:mvt_proba_poplevel}
\begin{aligned}
\omega_K( & \bt, \bSigma_1, \nu_2, \t,  \c)
\equiv
\Pr\{(\wh\bt, \wh\sigma_{1,1}, \ldots, \wh\sigma_{1,K})\trans \in C_K(\t,\c) \}\\
&=\Pr \left\{\bigcap_{k=1}^K\left(\frac{t_k\wh\sigma_{1,k}-c_k}{\sigma_{1,k}}-\frac{\theta_k}{\sigma_{1,k}}<\frac{\wh\theta_k-\theta_k}{\sigma_{1,k}}<\frac{c_k-t_k\wh\sigma_{1,k}}{\sigma_{1,k}}-\frac{\theta_k}{\sigma_{1,k}} \right)\right\}.
\end{aligned}
\ee
Note that
  $\wh\sigma_{1,k}$ involves $\nu_2$,
hence $\omega_K$ is a function of
$\nu_2$ as well. To control the size of the test at $\alpha_0$, we
restrict ourselves to the set $\calA_K^*(\alpha_0)$, where
\be
&&\calA_K^*(\alpha_0) \equiv\{(\t, \c):
\t \in \real_{\geq0}^K , \c \in \real_{>0}^K, 
\,
\omega_K \{ \blambda(\t,\c), \bSigma_1, \nu_2 , \t,  \c\}=
\alpha_0 \},\label{eq:adjustments_mvt}\\
\mbox{and}&& \blambda(\t,\c) \equiv ( \lambda_1, \ldots, \lambda_K )\trans \in \bm{\Lambda}(\bSigma_1, \nu_2, \t, \c ) \equiv \argsup_{\bt \notin (-c_0, c_0)^K } \;
\omega_K (\bt, \bSigma_1, \nu_2, \t, \c). \qquad\quad \label{eq:lambda_argsup}
\ee
Here $\blambda(\t,\c)$ also depends on $\bSigma_1$ and $\nu_2$,
  while we omit them for notational simplicity.
Note that the complement set of  $(-c_0, c_0)^K$ is the parameter
space under the null
hypothesis,
hence $\blambda(\t,\c)$ defined in \eqref{eq:lambda_argsup} leads to the
worst-case scenario of the rejection probability under H$_0$. In the
univariate case ($K=1$), it is obvious that  $\lambda(t,c)=\pm c_0$,
and we can simply consider 
$\lambda(t,c)=c_0$ for subsequent power analysis without loss of generality.
However, in the multivariate case ($K>1$), other than
$\blambda(\t,\c)$ being on the boundary 
of $(-c_0, c_0)^K$,
a closed form of $\blambda(\t,\c)$ is not attainable in general 
due to its complex relation to $\bSigma_1$.  See Appendix~\ref{appendix:mvt_example} for
an illustration of how $\blambda(\t,\c)$ is affected by the correlation in $\bSigma_1$.

In contrast to the univariate case, even for a fixed $\t$ in
\eqref{eq:adjustments_mvt},  there exists multiple $\c(\t)$'s, where $\c(\t) \equiv [c_1(\t), \ldots, c_K(\t)]\trans$, that
guarantee the size $\alpha_0$ simply because we have only a
  single requirement $\omega_K [ \blambda\{\t,\c(\t)\}, \bSigma_1, \nu_2 , \t,  \c(\t)]=
\alpha_0$ while we have $K>1$ degrees of freedom in $\c(\t)$. 
Indeed, if we start from a rejection region based on any $\{\t, \c(\t)\} \in
\calA_K^*(\alpha_0)$, then it is always possible to increase the rejection region
 across some dimensions while reducing it along some other dimensions
 to preserve the size of $\alpha_0$.
  In particular, highly unbalanced rejection region can occur, where
 some entries of $\c(\t)$ may exceed $c_0$ while others can be
 arbitrarily close to 0 (see Appendix~\ref{appendix:mvt_example} for an example), which may not be
 desirable in practice.
Thus, in addition to achieving an overall size of $\alpha_0$, we
further impose an equal marginal size requirement, i.e.,
$\omega \{ c_0,\sigma_{1,k}, \nu_2, t_k, c_k(\t) \}$ are identical
across all $k=1, \dots, K$. The marginal size can be viewed as the
size of the univariate test concerning only the $k$th component of
$\bt$.  This leads to the reduced set
\be \label{eq:adjustments_mvt_constr_size}
\begin{aligned}
\calA_K(\alpha_0) \equiv\{(\t, \c):
 \t & \in \real_{\geq0}^K , \c \in \real_{>0}^K, 
\,
\omega_K \{ \blambda(\t,\c), \bSigma_1, \nu_2 , \t,  \c\}=
\alpha_0 ,  \\
 \omega & (c_0, \sigma_{1,1}, \nu_2, t_1, c_1) = \dots = \omega ( c_0,
\sigma_{1,K}, \nu_2, t_K, c_K ) 
\} .
\end{aligned}
\ee
For a fixed $\t$, there is a unique solution $\c(\t)$ so that $\{\t,\c(\t)\} \in \calA_K(\alpha_0)$.
Building upon the univariate setting presented in
Section~\ref{sec:proposed:univariate}, one can expect that among all
potential $\{\t,\c(\t)\} \in \calA_K(\alpha_0)$,  setting $\t=\0$ would lead
to some desirable properties.  This prompts us to consider
$\c^* \equiv \c(\0)$ such that
\be
&&\omega_K\{\blambda(\0,\c^*), \bSigma_1, \nu_2 , \0,  \c^*\} =
\alpha_0,\label{eq:level}\\
\mbox{where}&&
    \blambda (\0, \c^*)
    \in \bm{\Lambda}(\bSigma_1, \nu_2, \0, \c^* )
    \equiv
    \argsup_{\bt \notin (-c_0, c_0)^K } \;
    \omega_K ( \bt, \bSigma_1, \nu_2 , \0,  \c^* ), \label{eq:mvt_ctost_sup}\\
\mbox{and}&& \omega (c_0, \sigma_{1,1}, \nu_2, 0, c_1^*) = \dots = \omega ( c_0,
\sigma_{1,K}, \nu_2, 0, c_K^* ) = \gamma. \label{eq:level_marginal}
\ee
Interestingly, we will always have $\gamma \geq \alpha_0$,
as the global size in \eqref{eq:level} cannot exceed the marginal sizes in \eqref{eq:level_marginal}, i.e.,
\be
\begin{aligned}
\alpha_0 = \omega_K\{\blambda(\0,\c^*), \bSigma_1, \nu_2 , \0,  \c^*\} &\leq 
\min_{k=1, \ldots, K} \,
\omega \{ \lambda_k(\0,\c^*), \sigma_{1,k}, \nu_2, 0, c_k^* \} \n \\
&\leq 
\omega \{ \lambda_h(\0,\c^*), \sigma_{1,h}, \nu_2, 0, c_h^* \}  = \gamma . 
\end{aligned}
\ee
The first inequality is a Fr\'echet inequality, the second inequality
holds since $\blambda(\0,\c^*)$ has at least one entry, say
$h$, 
at $\pm c_0$ to satisfy \eqref{eq:mvt_ctost_sup}, 
and the first and last equalities are respectively defined in
\eqref{eq:level} and \eqref{eq:level_marginal}. 
We denote the proposed adjusted TOST under the $(\0,\c^*)\in
  \calA_K(\alpha_0)$ choice as multivariate cTOST. 
Similarly to the univariate cTOST, for the multivariate cTOST,
  the probability of rejecting H$_0$,
$\omega_K(\bt, \bSigma_1, \nu_2, \0, \c^*)$, does not change if 
$\bSigma_1$ is known, given that we restrict to
the reduced set 
\eqref{eq:adjustments_mvt_constr_size}  as well. Furthermore, 
when $\bSigma_1$ is known, all size-$\alpha_0$ adjustments with equal
marginal sizes possess equal power under H$_1$ (see
Appendix~\ref{app:known_Sigma_case_mvt}).
  
Although the family of adjustments $\{\t,\c(\t)\} \in
  \calA_K(\alpha_0)$ guarantees the uniqueness of the test at any fixed
  $\t$, no choice of $\t$ can lead to a uniformly
  most powerful test in this family when $\bSigma_1$ is unknown, as demonstrated in
  Appendix~\ref{appendix:mvt_ctost}. 
To illustrate  the theoretical properties of multivariate cTOST, we
restrict our analysis to the 
scenario of $\hbtheta$ with independent homoskedastic components,
i.e.,~$\sigma_{1,1}=\ldots=\sigma_{1,K}$.
Here,  we show in
Proposition~\ref{pro:mvt_most_powerful_at_0} that
among all choices of $ \{\t,\c(\t)\} \in \calA_K(\alpha_0)$, 
the multivariate cTOST has the largest power when evaluated at
  the alternative $\btheta=\0$, 
which is of the greatest interest when assessing multivariate
bioequivalence.
\begin{pro} \label{pro:mvt_most_powerful_at_0}
At any given $\nu_2$, $\sigma_1$, consider
$\bSigma_1 = \sigma_1^2 \I_K$.
Then for all $\{\t, \c(\t)\} \in \calA_K(\alpha_0)$ in
\eqref{eq:adjustments_mvt_constr_size}, $\c^*$ defined in
\eqref{eq:level}-\eqref{eq:level_marginal} satisfies 
$
\omega_K \{ \0, \bSigma_1, \nu_2, \t, \c(\t) \}
\leq 
\omega_K \{ \0, \bSigma_1, \nu_2, \0, \c^* \} .
$
Hence, at the origin, the multivariate cTOST is the most powerful test
among all the tests defined by $(\t,\c)\in\calA_K(\alpha_0)$.
\end{pro}
The proof of Proposition~\ref{pro:mvt_most_powerful_at_0} is presented
in Appendix~\ref{appendix:mvt_ctost}. 
Proposition~\ref{pro:mvt_most_powerful_at_0} ensures that
under independence and homoskedasticity, among all (globally)
size-$\alpha_0$ tests with equal marginal sizes, the multivariate
cTOST is the most powerful test at the origin. 
We conjecture that this optimality property also holds under
heteroskedasticity and general correlation structures.  
Numerical results for bivariate settings support this conjecture (see
Appendix~\ref{appendix:mvt_ctost}). 
However, despite its optimality at $\btheta=\0$, the multivariate
cTOST may become less powerful in some regions of the alternative
space, such as the one close to $\btheta = c_0\1_K$ (where $\1_K$ is a length $K$ vector of ones). 
This result is demonstrated in Appendix~\ref{appendix:mvt_ctost} under
the same assumptions of Proposition~\ref{pro:mvt_most_powerful_at_0}.
Furthermore, in Appendix~\ref{appendix:mvt_ctost_vs_atost} we illustrate
the superior operating characteristics of multivariate cTOST versus multivariate $\alpha$-TOST \citep{boulaguiem2025mvt}, 
which belongs to the set $\calA_K^*(\alpha_0)$ in \eqref{eq:adjustments_mvt}.

The construction of a feasible finite-sample
adjustment $\wh \c^*$ for multivariate cTOST, where population
quantities in \eqref{eq:level}-\eqref{eq:level_marginal} are replaced
by their empirical counterparts, follows similar procedures to that
described in Section~\ref{sec:proposed:univariate} for the univariate
cTOST. 
Operationally, $\wh\c^*$ can be efficiently computed by leveraging the
algorithm for the univariate cTOST described in
Appendix~\ref{appendix:algo_ctost}.  
A formal description of our algorithm is provided in
Appendix~\ref{appendix:algo_ctost_mvt}.
Similarly to the univariate framework, if $\lvert \wh 
\theta_k \rvert < \wh c^{\;*}_k$, for all $k=1, \ldots, K$,
H$_0$ is rejected at the significance level $\alpha_0$.

\subsection{Further Small-Sample Refinements}
\label{sec:finite_refinement}

For reasonable sample sizes, 
cTOST in 
\eqref{eq:whc0} is very accurate at controlling the size of the
test, as illustrated in the numerical results of
Section~\ref{sec:simulation}. 
Specifically, for most practical applications, where $\nu_2$ is not too small,
it effectively
maintains the size at $\alpha_0$.
When sample size is extremely small,
cTOST may not adequately control the test size.
Given the critical importance of controlling the test size in
bioequivalence studies, as it relates to the ``consumer'' or
``regulator'' risk \citep{patterson2017bioequivalence}, we further
refine cTOST so that it maintains the test size even
under very limited amount of data.

Specifically, we replace $\alpha_0$ with a ``calibrated'' level $\alpha_c \leq
\alpha_0$ to further adjust $\wh c(0)$ in~\eqref{eq:whc0}. 
The empirical size associated to $\wh c(0, \hsigma_1, \alpha_0) \equiv
\wh c(0)$ defined in \eqref{eq:whc0}  can be expressed as  
$ 
\omega \{ c_0, \sigma_1, \nu_2, 0, \wh c(0, \wh\sigma_1, \alpha_0) \}
= \Pr\big\{\vert\wh{\theta}\vert  \leq  \wh c(0, \wh\sigma_1,
\alpha_0)\big\lvert \theta=c_0, \sigma_1 = \sigma_1 \big\},$
which incorporates the uncertainty associated with the estimation of $\sigma_1$.
This is in contrast to the theoretical size based on a plug-in of $\hsigma_1$, that is,
$
\omega \{ c_0, \wh\sigma_1, \nu_2, 0, \wh c(0, \wh\sigma_1, \alpha_0)
\} = \Pr\{ |\wh \theta| <  \wh c(0, \wh\sigma_1, \alpha_0) |
\theta=c_0, \sigma_1 = \wh\sigma_1\}, 
$
used to solve the matching in \eqref{eq:whc0}.
Therefore, a calibrated level $\alpha_c$ 
can be defined as 
\be \label{eq:univ_ctost_corr}
\alpha_c \equiv \argzero_{\gamma  \in (0, \alpha_0]}
    E[ \omega \{ c_0, \hsigma_1, \nu_2,
      0, \wh c(0, \hsigma_1^{\star}, \gamma) \}  ] - \alpha_0 , 
\ee
where 
$\nu_2 \wh\sigma^{\star2}_{1} / \wh\sigma_1^2 \sim \chi^2_{\nu_2}$
conditionally on $\wh\sigma_1$. 
A simple iterative approach can be used to compute $\alpha_c$, where
at each iteration $j$, with $j \in \mathbb{N}$, we update
$\alpha_c^{(j+1)} = \alpha_0 + \alpha_c^{(j)} - E[ \omega \{ c_0,
\hsigma_1, \nu_2, 0, \wh c(0, \hsigma_1^{\star}, \alpha_c^{(j)}) \}
]$, initializing the procedure at $\alpha_c^{(0)} = \alpha_0$.
In practice, a single iteration is often
sufficient to ensure a reliable test size.
We then let $\wt c(0)$ 
denote a  refined version of $\wh c(0)$ in \eqref{eq:whc0} defined as:
\be \label{eq:univ_ctost_calib}
\wt c(0) \equiv \wh{c} (0, \wh\sigma_1, \alpha_c) =  \argzero_{ c \in \real_{>0} }\ \{ \omega(c_0,
\wh\sigma_1, \nu_2, 0, c)-\alpha_c\} .
\ee
Several approaches can be used to approximate the
  expectation in \eqref{eq:univ_ctost_corr}, such as the
  bootstrap. 
For given $\alpha_0$ and $c_0$, a precomputed table
containing approximations of the expectation in
\eqref{eq:univ_ctost_corr}, based on a grid of $\sigma_1$ and $\nu_2$
values, can be stored in software allowing for efficient computation
of $\widetilde \alpha_c$ (i.e.,~the empirical counterpart of $\alpha_c$) without additional computational overhead.  
For the customary equivalence margins and significance level that are
used by regulatory agencies in bioequivalence studies (i.e.,~$c_0=\log(1.25) \approx 0.223$
and~$\alpha_0 = 0.05$), 
\if0\blind{
we provide the resulting table as a
part of our software 
implementation in the \texttt{cTOST} R package. 
} \fi
\if1\blind{
we provide the resulting table as a
part of our software 
implementation.  
} \fi
Additional details on the small-sample refinement, along with
simulation results demonstrating its effectiveness for very small
sample sizes (e.g.,~$\nu_2=5$), are presented in
Appendix~\ref{appendix:sim_univ_main}. 
When necessary, a similar approach can be applied to 
multivariate cTOST adjustments $\wh \c^*$ described in
Section~\ref{sec:proposal_multiv}.

\section{Simulation Studies}
\label{sec:simulation}

We present the results of extensive univariate and
  multivariate simulation studies to compare the performance of
  various methods.

\subsection{Univariate Setting}
\label{sec:simulation_univ}

For univariate simulations, we consider the canonical
  form defined in \eqref{eq:mvt_canon} for $K=1$. 
We vary $\sigma_1$ over $1000$ equally-spaced values in the
interval $[0.01, 0.2]$ with degrees of freedom
$\nu_2 \in \{ 20, 40, 80 \}$, at the nominal significance level
$\alpha_0 = 0.05$ and equivalence bound $c_0 = \log(1.25)$. 
Four methods are compared: the standard TOST, the $\alpha$-TOST, the
cTOST based on $\wh c(0)$ in \eqref{eq:whc0}, and its 
refined counterpart (cTOST*) based on $\wt{c}(0)$ in
\eqref{eq:univ_ctost_calib}. The study is based on $10^5$ Monte Carlo
replications.

\begin{figure}[t!]
        \vskip -1.3cm
        \centering
        \includegraphics[width=1\textwidth]{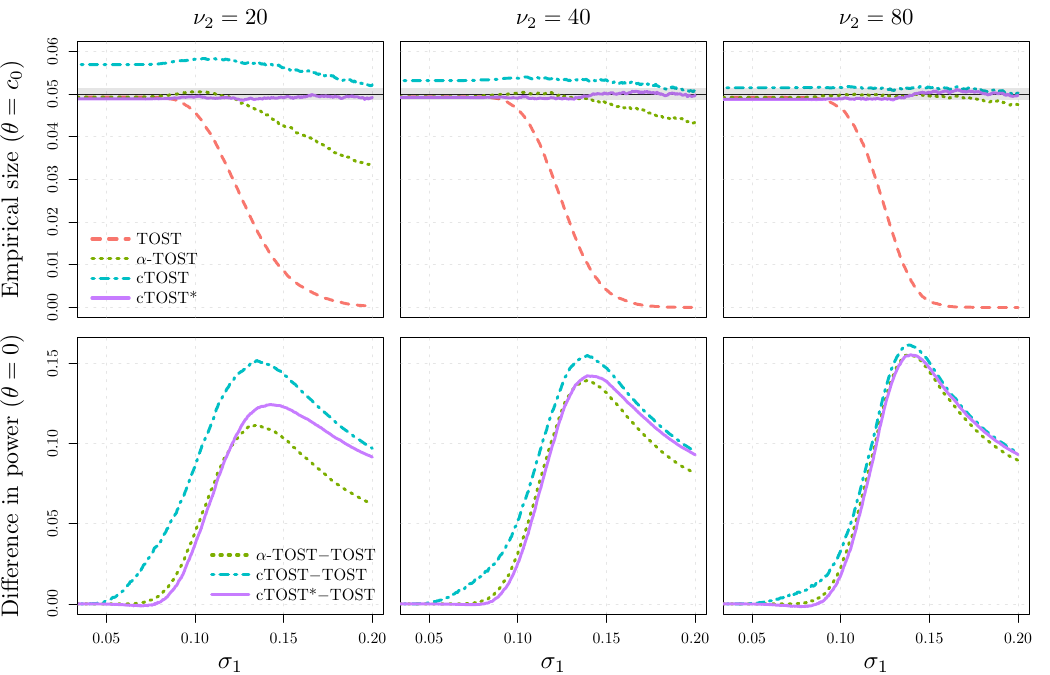}
        \caption{Simulation results comparing the performance of TOST, $\alpha$-TOST, cTOST and cTOST*.
        Top row: Empirical size. Bottom row: Difference in power at $\theta=0$ with respect to TOST.}
        \label{fig:curves_sigma_univ}
\end{figure}

The empirical size of the different methods is shown in
the first row of Figure~\ref{fig:curves_sigma_univ}, while their power is
evaluated by setting $\theta = 0$, and is in
the second row, where we report
the power difference between 
$\alpha$-TOST, cTOST, and cTOST* relative to the standard
TOST.
These results demonstrate that both the cTOST$^*$ and cTOST can
substantially improve the power of the testing procedure compared to
the TOST, and to a lesser extent, the $\alpha$-TOST.
While cTOST remains most powerful across all settings, it
exhibits a slightly inflated size for small $\nu_2$'s.
On the other hand, the empirical size of cTOST* remains consistently
close to $\alpha_0$, staying within the simulation error bounds of
this numerical experiment (represented as gray regions around the
$\alpha_0$ value, denoted by a black solid line), while still
  having generally larger power than $\alpha$-TOST and TOST.  

Additional simulation results are presented in
Appendix~\ref{appendix:sim_univ_main}, where we more closely study the
power and size of various test procedures, examine a wider range of 
$\theta$ values, and consider scenarios with fewer degrees of freedom
$\nu_2$.  
Across all settings, the performance remains consistent with the
  results presented in 
Figure~\ref{fig:curves_sigma_univ}.

\subsection{Multivariate Setting}
\label{sec:simulation_multiv}

For multivariate simulations, we consider the canonical form in~\eqref{eq:mvt_canon} for $K>1$.
To facilitate a fair comparison across test procedures, we consider 
$ 
\widehat{\bm{\theta}} \sim \mathcal{N}_K \left\{ \kappa \blambda ,\bm{\Sigma}_1 \right \},
$
where the vector $\blambda$ defined in \eqref{eq:lambda_argsup}
denotes the coordinates in the parameter space under the null
hypothesis that determine the method-specific test size.
We then vary the parameter $\kappa$ across 30 equally-spaced values in
the interval  $ [ 0, 1.2 ] $. Therefore, while the test size is
evaluated at $\kappa=1$, the probabilities of declaring equivalence
under the alternative and null hypotheses are respectively evaluated
at $\kappa \in [0,1)$ and $\kappa \in (1,1.2]$. 
For the covariance matrix $\bm{\Sigma}_1$, we consider $\Sigma_{1,jk}
= \rho \sigma_{1,j} \sigma_{1,k}$ for $j \neq k$ and $ \Sigma_{1,jk} =
\sigma_{1,j}^2 $ otherwise, where $ \lvert \rho \rvert < 1$ denotes
the pairwise correlation.

\begin{figure}[t!]
    \vskip -1.75cm
    \centering
    \includegraphics[width=1\textwidth]{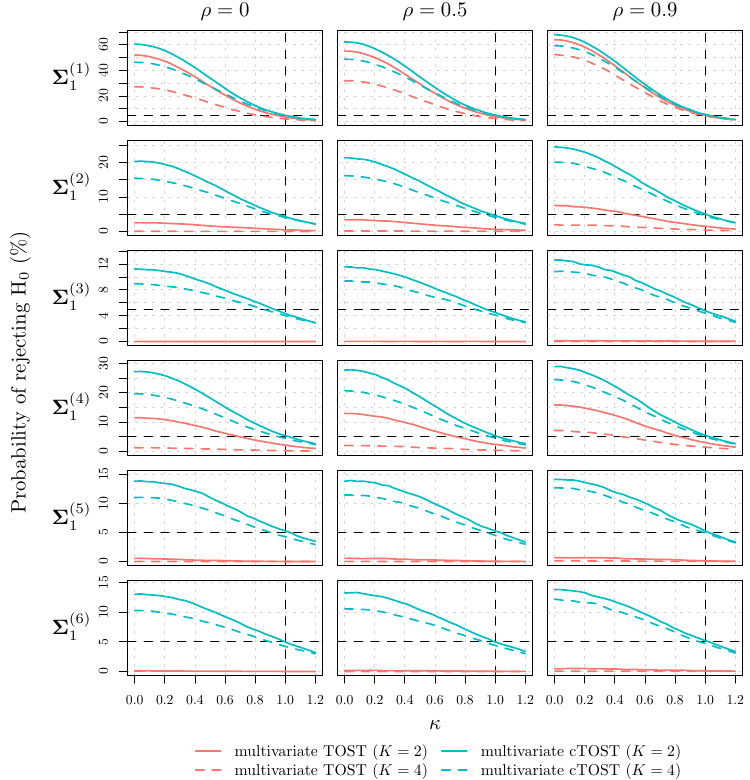}
    \caption{Probability of rejecting the null hypothesis as a function of $\kappa$, for the multivariate TOST and the multivariate cTOST, across different levels of correlation $\rho$ (columns) and covariance matrices (rows), for the bivariate (solid lines) and 4-variate (dashed lines) settings.}
  \label{fig:mvt_power_sim}
\end{figure}

We consider two dimensions $K\in\{ 2, 4\}$, three correlation levels $\rho\in\left\{0, 0.5, 0.9\right\}$, and six configurations of the covariance matrix $\bm{\Sigma}_1 \in \{ \bm{\Sigma}_1^{(1)}, \ldots, \bm{\Sigma}_1^{(6)} \} $. 
In the bivariate setting, the six configurations are
$ 
(\sigma_1, \sigma_2) \in \{ (0.08, 0.08), (0.12, 0.12), (0.16, 0.16), \\ (0.08, 0.12), (0.08, 0.16), (0.12, 0.16) \}.
$
In the 4-variate setting,  we set $\sigma_1=\sigma_2$ and
$\sigma_3=\sigma_4$, where each pair takes the same six sets of
  values. 
The study is based on $5 \times 10^4$ Monte Carlo replications.

Figure~\ref{fig:mvt_power_sim} compares the performance of the
multivariate TOST and the multivariate cTOST -- no further refinement
was necessary for cTOST due to its good performance in these settings.
Similarly to the univariate case presented in
Section~\ref{sec:simulation_univ}, the multivariate cTOST is 
more powerful than TOST across all simulation settings and,
importantly, its empirical size remains very close to the nominal
level $\alpha_0$. 
As expected, it provides greater power gains in the presence of lower
levels of correlation, higher dimensionality, and larger variances.
The multivariate $\alpha$-TOST performs
better than 
 the multivariate TOST, but it remains less powerful than cTOST across all settings 
 (see Appendix~\ref{appendix:sim_mvt_main}). 
Furthermore, the multivariate cTOST offers improved
computational efficiency and stability compared to the
$\alpha$-TOST. The latter may also lead to fairly uneven rejection regions
(i.e.,~leading to very different marginal sizes), failing to reflect
the practical situations of interest to scientists, as 
illustrated in the case study in
Section~\ref{sec:application}.

\section{Case Study: Cutaneous Bioequivalence}
\label{sec:application}

\citet{quartier2019cutaneous} conducted a study on the
  cutaneous bioequivalence between two topical cream products: a
brand-name cream (\textit{Pevaryl}) and an approved
  generic formulation (\textit{Mylan}) containing econazole nitrate
  (ECZ), an antifungal medication commonly utilized in the treatment
  of skin infections.
The evaluation of putative bioequivalence relies on analyzing the
cutaneous biodistribution profile of ECZ following the application of
\textit{Pevaryl} versus \textit{Mylan}. This entails
quantifying ECZ concentrations at different skin depths, ranging from
the surface to a depth of approximately 800 $\mu m$. The similarity in
biodistribution profiles -- indicative of comparable drug
concentration at each skin depth -- is used to evaluate their bioequivalence.
The dataset consists of $n=12$ independent pairs of comparable
porcine skin samples on which ECZ deposition measurements were
collected after applying the two topical products. 
The sample size is heavily limited by the complexity of the
experimental protocol, leading to $\nu_1 = 12$ and $\nu_2 = 11$.
Measurements at 21 different skin depths 
recorded in \citet{quartier2019cutaneous}
were grouped to
obtain $K=4$ anatomically relevant regions: 
\textit{stratum corneum} (0–20 $\mu m$; $\htheta_1 \approx 0.0976$, $\hsigma_{1,1} \approx 0.3305$),
\textit{viable epidermis} (20–160 $\mu m$; $\htheta_2 \approx 0.0726$, $\hsigma_{1,2} \approx 0.1708$), \textit{upper dermis}
(160–400 $\mu m$; $\htheta_3\approx0.0022$, $\hsigma_{1,3} \approx 0.2233$) and \textit{lower dermis} (400–800 $\mu m$; $\htheta_4 \approx 0.0733$, $\hsigma_{1,4} \approx 0.1853$).

\begin{table}[t!]
\vskip -1.2cm
\hspace*{-0.6cm}
\footnotesize
\centering
\begin{tabular}{lccccc}
  \toprule
  Multivariate
 & \textit{Stratum corneum} & 
 \textit{Viable epidermis} & 
 \textit{Upper dermis} & 
 \textit{Lower dermis} & 
  Bioequivalence \\
 method & (0–20 $\mu m$) & 
 (20–160 $\mu m$) & 
 (160–400 $\mu m$) & 
 (400–800 $\mu m$) & 
 declaration \\ 
  \midrule
TOST & [-0.496, 0.691] & [-0.234, 0.379] & [-0.399, 0.403] & [-0.259, 0.406] & {\color{red!90!black}\xmark} \\ 
  $\alpha$-TOST  & [-0.034, 0.229] & [\,\,0.005, 0.141] & [-0.087, 0.091] & [\,\,0.001, 0.147] & {\color{red!90!black}\xmark} \\ 
  cTOST &[\,\,0.012, 0.184] & [-0.027, 0.172] & [-0.101, 0.106] & [-0.029, 0.175] & {\color{green!70!black}\cmark} \\ 
   \bottomrule
\end{tabular}
\caption{Confidence intervals obtained using the multivariate TOST,
 $\alpha$-TOST, and cTOST for the
  four-dimensional case study related to econazole nitrate deposition
  in porcine skin. The rejection region is $(-c_0,c_0)$ with $c_0 =
  \log(1.25) \approx 0.223$ and the nominal significance level is
  $\alpha_0=5\%$. Bioequivalence is declared by a method when all intervals are within $(-c_0, c_0)$.
} 
\label{tab:app_ci} 
\end{table}

The results of multivariate bioequivalence assessment using the
multivariate TOST, multivariate $\alpha$-TOST, and multivariate cTOST
are presented in Table~\ref{tab:app_ci}. 
As customary in this literature, we set $c_0=\log(1.25)$ and
$\alpha_0=5\%$. 
Based on the conservative behavior of multivariate cTOST reported in
an emulation study based on these data (not provided here), no further
refinement was pursued.
We present marginal confidence intervals for each of the four  
biodistribution profiles, as well as an indication of whether the test
procedure declares bioequivalence (i.e.,~whether each marginal
confidence interval is contained in the equivalence margins according
to the IIP). 
Since $c_0 > \wh c^{\;*}_k$ for all $k=1,\ldots,K$, cTOST also allows
for an IIP-based interpretation by considering the intervals $\wh
\theta_k \pm \left\{c_0 -  \wh c^{\;*}_k \right\}$ (see
Appendix~\ref{appendix:algo_ctost_mvt}).
Both the multivariate TOST and multivariate $\alpha$-TOST fail to
declare bioequivalence in the considered application.  
In contrast, the  multivariate cTOST detects
bioequivalence between the two topical formulations. 
\begin{figure}[!t]
\vskip -1.75cm
\hspace*{-1.5cm}
\centering
\includegraphics[width=1.2\textwidth]{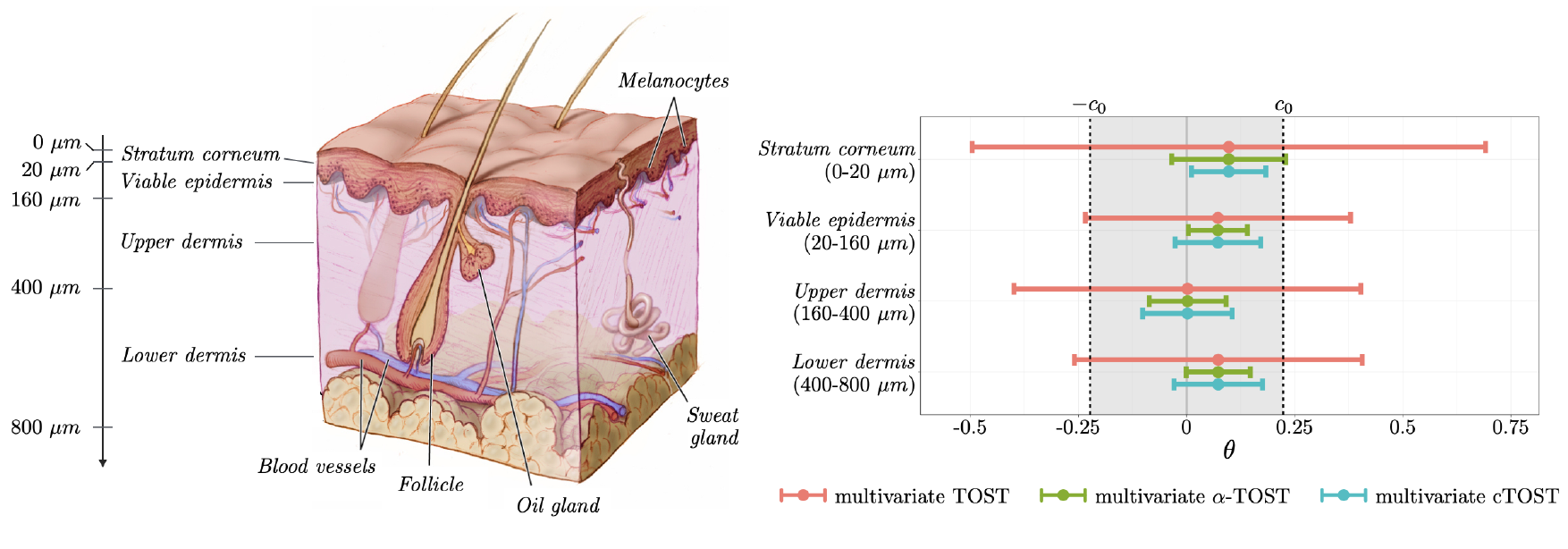} 
\caption{Left: cross-sectional diagram of the skin highlighting its
  key structures. Right: Representation of the confidence intervals
  obtained by the multivariate TOST, $\alpha$-TOST, and cTOST as
  reported in Table~\ref{tab:app_ci}  for the four-dimensional case
  study related to econazole nitrate deposition in porcine skin. 
  Bioequivalence is declared by a method when all intervals are
    within $(-c_0, c_0)$. 
\label{fig:app_results}}
\end{figure}
This distinction is further illustrated in
Figure~\ref{fig:app_results}, which depicts the bioequivalence
assessment problem in terms of equivalence margins and marginal
confidence intervals.  
In particular, while the TOST becomes extremely conservative, the
$\alpha$-TOST leads to an improvement in terms of power by producing
shorter confidence intervals. However, the $\alpha$-TOST suffers not
only from its marginal lack of power compared to the cTOST but also as
it corrects all dimensions in the same manner. Instead, the cTOST
leads to more balanced rejection regions as highlighted by the length
of its confidence intervals, where its longest confidence interval
remains shorter than the ones provided by competing methods.

\section{Final Remarks}
\label{sec:conclusion}

In this article, we presented a novel cTOST procedure in the context of multivariate average equivalence testing.
The univariate cTOST
is the uniformly most powerful test under known
variance, and remains  the optimal adjustment when
the variance is estimated. In small samples, we propose a 
 refinement for cTOST, 
which can be
computed easily, and can ensure that the test size remains
close to the nominal significance level. 
The univariate refinement similarly applies to the multivariate
setting. 
We find the multivariate setting much more complex,
  with no single adjustment remaining uniformly more powerful over
  the entire alternative space even under a known covariance structure.
To complicate matters further, for any fixed parameter vector in the
alternative space, the most powerful adjustments often correspond to
rejection regions that are highly unbalanced across the $K$ dimensions
(i.e.,~marginal rejection regions close to zero in some dimensions and
potentially unbounded for others), failing to reflect the practical
situations of interest to scientists.
Therefore, we constrain the marginal size of each univariate test to a
constant and propose multivariate cTOST, which
retains certain optimality, is computationally
simple,  and yields rejection regions that align with
practical needs in scientific applications.

By improving the efficiency and reliability of bioequivalence studies,
which play a pivotal role in biopharmaceutical development
\citep{patterson2017bioequivalence}, the cTOST method can facilitate the
approval process for generic medical products. This is particularly
important for locally acting drugs, where current methodological
constraints often limit the development of generic products
\citep{miranda2018bioequivalence,miranda2018bioequivalence_p2}. 
These advancements align with the World Health Organization’s goal of ensuring access to safe, effective, high-quality, and affordable medicines, advancing the vision of Universal Health Coverage, particularly in low- and middle-income countries where affordability and availability remain key obstacles \citep{wirtz2017essential}.
The broad applicability of our methodology extends beyond traditional
bioequivalence studies to general equivalence tests across various
disciplines  (\citealt{Lakens:18}, 
\citealt{moore2022engineering}, \citealt{aggarwal20232}). 
The cTOST method specifically addresses common challenges across these
domains, such as reduced sample sizes and/or the presence of
multivariate outcome measures.  By achieving larger power while
maintaining  type I error rates, cTOST 
provides practical solutions for researchers and practitioners
across different fields.

\begingroup
    {\small
    \singlespacing
    \bibliography{refs.bib}
    }
\endgroup

\cleardoublepage
\bigskip
\begin{center}
{\large\bf SUPPLEMENTARY MATERIAL}
\end{center}


\appendix
\label{sec:app}

\renewcommand{\thetable}{A.\arabic{table}}
\setcounter{table}{0}
\renewcommand{\theequation}{A.\arabic{equation}}
\setcounter{equation}{0}
\renewcommand{\thefigure}{A.\arabic{figure}}
\setcounter{figure}{0}

\section{Equivalence of all adjustments in  $\calA(\alpha_0)$ when $\sigma_1$ is known}
\label{appendix:power_known_sigma}

When $\sigma_1$ is known, the canonical form of the univariate average equivalence problem described in \eqref{eq:mvt_canon} for $K=1$ reduces to
\bse
    \widehat{\theta} \sim \mathcal{N}\left(\theta, \sigma_1^2\right) .
\ese
The TOST is thus based on the two following test statistics:
\bse
Z_L \equiv \frac{\widehat{\theta}+c}{\sigma_1} \quad\text{and}\quad Z_U \equiv \frac{\widehat{\theta}-c}{\sigma_1},
\ese
where $Z_L$ tests for $\text{H}_{01}: \theta\leq -c$ versus $\text{H}_{11}: \theta> -c$, and $Z_U$ tests for $\text{H}_{02}: \theta \geq c$ versus $\text{H}_{12}: \theta<c$. 
At a significance level $\alpha$, we reject $\text{H}_{0}\equiv\text{H}_{01}\cup\text{H}_{02}$ in favor of $\text{H}_{1}\equiv\text{H}_{11}\cap\text{H}_{12}$  if both tests simultaneously reject their marginal null hypotheses, that is, if
\bse
     Z_L \geq z_{\alpha} \;\;\; \text{{and}} \quad Z_U \leq -z_{\alpha},
\ese
where $z_{\alpha}$ denotes the upper $\alpha$ quantile of a normal distribution.  Using 
$\omega(c_0, \sigma_1, \nu_2, 0, c)$ defined after
\eqref{eq:xtost_matching} and letting the size of the test to be $\alpha_0$, we have
\bse 
\omega(c_0, \sigma_1, \infty, z_\alpha, c )
=
\Phi\left(\frac{c-c_0}{\sigma_1}-z_{\alpha}\right)-\Phi\left(-\frac{c+c_0}{\sigma_1}+z_{\alpha}\right)
= \alpha_0 .
\ese

This allows us to determine a unique $c(z_\alpha)$ for any
subjectively selected $z_\alpha$, i.e.,~for any $z\ge0$, $c(z)$ is such that 
\bse
\omega \{ c_0, \sigma_1, \infty, z,  c(z) \} 
=
\Phi\left\{\frac{c(z)-c_0}{\sigma_1}-z\right\}-\Phi\left\{-\frac{c(z)+c_0}{\sigma_1}+z\right\}=\alpha_0.
\ese
The power of such a test is 
\bse
\omega \{ \theta, \sigma_1, \infty,  z, c(z) \}  
=
\Phi\left\{\frac{c(z)-\theta}{\sigma_1}-z\right\}-\Phi\left\{-\frac{c(z)+\theta}{\sigma_1}+z\right\}.
\ese
Because
\bse
\alpha_0=
\Phi\left[-\frac{c_0}{\sigma_1}+\left\{\frac{c(z)}{\sigma_1}-z\right\}
\right]
-\Phi\left[-\frac{c_0}{\sigma_1}-\left\{\frac{c(z)}{\sigma_1}-z\right\} \right],
\ese
hence $\frac{c(z)}{\sigma_1}-z$ is unique regardless of the choice of
$z$. 
Thus, 
\bse
\omega\{\theta, \sigma_1,  z, \infty, c(z)\} 
=
\Phi\left[ -\frac{\theta}{\sigma_1}+\left\{\frac{c(z)}{\sigma_1}-z\right\}
\right]
-\Phi\left[-\frac{\theta}{\sigma_1}-\left\{ \frac{c(z)}{\sigma_1}-z \right\} \right],
\ese
does not change as $z$ changes. This means regardless of the choice of
$z$, any test at size $\alpha_0$ has the same power.

\begin{figure}[ht]
    \centering
    \includegraphics[width=0.8\textwidth]{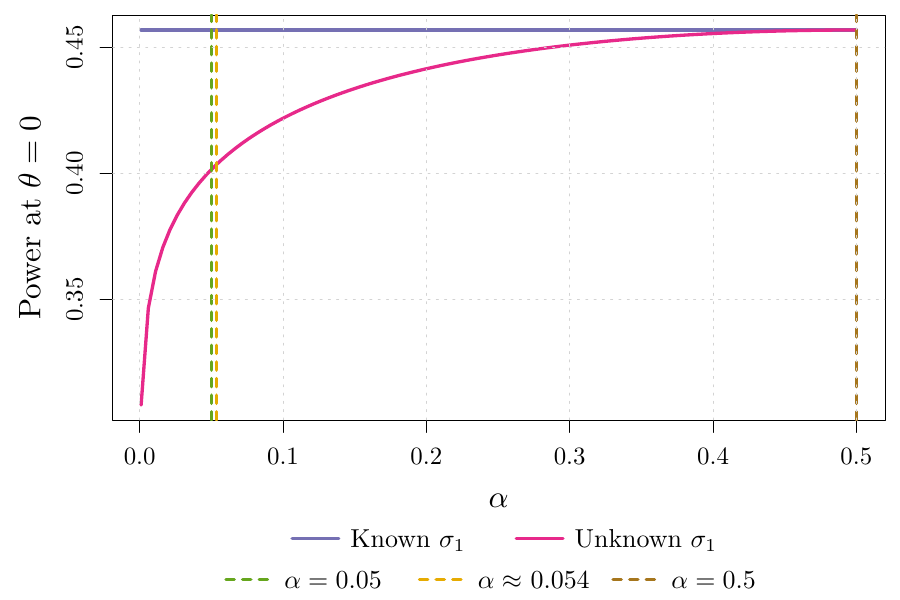}
    \caption{Power of adjusted TOST procedures at $\theta=0$ as a
      function of $\alpha$  when $\sigma_1$ is known (flat blue line)
      and unknown (increasing pink curve). 
    Dashed vertical lines represent the increasing $\alpha$ values
    associated to the $\delta$-TOST ($\alpha=0.05$),
    $\alpha$-TOST ($\alpha=0.054$), and cTOST ($\alpha=0.5$), respectively.}
  \label{fig:power_adjusted_tosts}
\end{figure}

As an illustration, Figure~\ref{fig:power_adjusted_tosts}
compares the power at $\theta=0$ as a function of $\alpha$
in both the unknown and known $\sigma_1$ cases.
We set $\sigma_1 = 0.1$, $\nu_2=15$, $c_0=\log(1.25)$, and
$\alpha_0=0.05$. 
The pink curve represents the power $\omega
\{\theta, \sigma_1, \nu_2, \alpha,\nu_2,
c(t_{\alpha,\nu_2},\nu_2)\}$ when $\sigma_1$ is unknown. This curve
is upper bounded by the constant blue line, which represents the power
when $\sigma_1$ is known.
The dashed vertical lines correspond to values of $\alpha$
associated with the $\delta$-TOST, $\alpha$-TOST, and cTOST,
respectively. This shows that setting $\alpha$ to $1/2$ results in the
most powerful adjustment when $\sigma_1$ is unknown.

\section{Proof of Theorem~\ref{thm:most_powerful}}
\label{appendix:_power_xtost_univ}

Consider the test 
\bse
\text{H}_0: \mu\ge c_0/\sigma_1 
\mbox{ or } \mu\le c_0/\sigma_1 
, \,\,\,\, \text{H}_1:-c_0/\sigma_1<\mu<c_0/\sigma_1,
\ese
using a single observation $X=\wh\theta/\sigma_1\sim
N(\mu, 1)$ with known $\sigma_1$.
Below, we use Theorem 3.5 in \citet{li2019graduate} to prove that
the test with rejection
 region  $(-c(0) /\sigma_1,c(0) /\sigma_1)$ is the most powerful test at level $\alpha_0$
 in this case. 
Define the decision function 
$\wt\phi(X)=1$ is $X\in (-c(0)/\sigma_1,c(0) /\sigma_1)$ and $\wt\phi(X)=0$ if $X\in
(\infty,-c(0) /\sigma_1]\cup[c(0) /\sigma_1,\infty)$.
Then we can check that $\wt\phi(X)$ satisfies Lemma 3.9 when we set
$f(x)$ to be
the normal pdf with mean $-c_0/\sigma_1$ and variance 1, 
$h(x)$ to be
the normal pdf with mean $c_0/\sigma_1$ and variance 1, 
$Y(x)=x$, $t_1=-c(0) /\sigma_1$, $t_2=c(0) /\sigma_1$,
$\gamma_1=\gamma_2=0$, and
$\alpha=\alpha_0$. In particular, we can verify that
\bse
E_{\mu=-c_0/\sigma_1}\wt\phi(X)=
\Phi\left(\frac{c(0)+c_0}{\sigma_1}\right)
-\Phi\left(\frac{-c(0)+c_0}{\sigma_1}\right)
=\alpha_0
\ese
due to the definition of $c(0)$.
Likewise, 
\bse
E_{\mu=c_0/\sigma_1}\wt\phi(X)&=&
\Phi\left(\frac{c(0)-c_0}{\sigma_1}\right)
-\Phi\left(\frac{-c(0)-c_0}{\sigma_1}\right)\\
&=&
\Phi\left(\frac{c(0)+c_0}{\sigma_1}\right)
-\Phi\left(\frac{-c(0)+c_0}{\sigma_1}\right) \\
&=&\alpha_0.
\ese
Now result (2) in Theorem 3.5 implies the test is the most powerful, where the
power at $\theta/\sigma_1$ is 
\bse
\Phi\left(\frac{c(0)+\theta}{\sigma_1}\right)
-\Phi\left(\frac{-c(0)+\theta}{\sigma_1}\right) 
=\omega \{\theta, \sigma_1, \infty, 0, c(0) \}.
\ese
Now consider the test of the same hypotheses with rejection region
\bse
\wh\theta/\sigma_1\in ( t-c(t,\nu_2) /\sigma_1,c(t,\nu_2) /\sigma_1-t ),
\ese
 where
$c(t,\nu_2)$ satisfies
\bse
\omega\{c_0, \sigma_1, \nu_2,t, c(t,\nu_2)\}=\alpha_0. 
\ese
This is a test with the same level, hence it is less powerful, thus, 
$\omega (\theta, \sigma_1, \nu_2, t, c(t,\nu_2))
\le \omega \{\theta, \sigma_1, \infty, 0, c(0) \} = \omega \{\theta, \sigma_1, \nu_2, 0, c(0)\}$.
\qed

\section{Proof of Proposition~\ref{pro:asymp}}
\label{appendix:pro_asymp}

\begin{proof}[Proof of result~1 from Proposition~\ref{pro:asymp}]
First of all, $\nu_2\wh\sigma^2/\sigma^2$ has a chi-squared distribution
with mean $\nu_2$, variance $2\nu_2$, hence
$\wh\sigma^2/\sigma^2=1+O_p(\nu_2^{-1/2})$, thus 
$\wh\sigma_1/\sigma_1=1+O_p(\nu_2^{-1/2})$.
Next, from
$
\Phi\left(\frac{c_0+c(0)}{\sigma_1}\right)-
\Phi\left(\frac{c_0-c(0)}{\sigma_1}\right)=\alpha_0$
and $\{c_0+c(0)\}/\sigma_1\to\infty$, 
we have $c(0)<c_0$ for sufficiently large $\nu_1$. For sufficiently
large $\nu_1$, we have
\bse
1-\alpha_0>\Phi\left(\frac{c_0-c(0)}{\sigma_1}\right)
=
\Phi\left(\frac{c_0+c(0)}{\sigma_1}\right)-
\alpha_0
>1-2\alpha_0,
\ese
thus, $0<\{c_0-c(0)\}/\sigma_1\asymp1$. This implies
\be\label{eq:bound1}
c_1\sigma_1<c(0)-c_0<c_2\sigma_1,
\ee
 where
$c_1\equiv\Phi^{-1}(1-\alpha_0), c_2\equiv\Phi^{-1}(1-2\alpha_0)$ are two
positive constants. 

From a Taylor expansion, we get
\bse
\alpha_0&=&
\Phi\left(\frac{
c_0+ \wh c(0)}{\wh\sigma_1} \right) 
        -  \Phi\left(\frac{c_0 -\wh c(0) }{\wh\sigma_1}
        \right)\\
&=&\Phi\left(\frac{
c_0+  c(0)}{\sigma_1} \right) 
        -  \Phi\left(\frac{c_0 -c(0) }{\sigma_1}
        \right)\\
&&+\left\{\phi\left(\frac{
c_0+  c(0)}{\sigma_1} \right)+
\phi\left(\frac{
c(0)-  c_0}{\sigma_1} \right)\right\}\frac{\wh c(0)-c(0)}{\sigma_1}\\
&&
-\left\{\phi\left(\frac{
c_0+  c(0)}{\sigma_1} \right)\frac{c(0)+c_0}{\sigma_1}
+\phi\left(\frac{c(0)-  c_0}{\sigma_1}
\right)\frac{c(0)-c_0}{\sigma_1}\right\}\frac{\wh\sigma_1-\sigma_1}{\sigma_1}+r_1\\
&=&\alpha_0+
\phi\left(\frac{
c(0)-  c_0}{\sigma_1} \right)\frac{\wh c(0)-c(0)}{\sigma_1}
-\phi\left(\frac{c(0)-  c_0}{\sigma_1}
\right)\frac{c(0)-c_0}{\sigma_1}\frac{\wh\sigma_1-\sigma_1}{\sigma_1}+r_2\\
&=&\alpha_0+
\phi\left(\frac{
c(0)-  c_0}{\sigma_1} \right)\frac{\wh c(0)-c(0)}{\sigma_1}
+O_p(\nu_2^{-1/2}),
\ese
where $r_1, r_2$ are the residual terms that are of higher order. This leads to
$\wh c(0)-c(0)=O_p(\sigma_1\nu_2^{-1/2})$. This in combination with $c(0)=c_0
+O_p(\sigma_1)\asymp1$ leads to
$\wh c(0)-c(0)=o_p\{c(0)\}$.
\end{proof}

\begin{proof}[Proof of result~2 from Proposition~\ref{pro:asymp}]
First, from
\bse
\alpha_0&=&
\int_0^{c(t,\nu_2)/t} 
\left[\Phi\left(\frac{
c_0+  c(t,\nu_2) - x t}{\sigma_1} \right) 
        -  \Phi\left(\frac{c_0 -c(t,\nu_2) + x  t }{\sigma_1} \right)
          \right]
          f_W ( x\lvert \sigma_1,\nu_2)
        dx\\
&<&\int_0^{\infty}
\left[\Phi\left(\frac{
c_0+  c(t,\nu_2) - x  t}{\sigma_1} \right) 
        -  \Phi\left(\frac{c_0 -c(t,\nu_2) + x  t }{\sigma_1} \right)
          \right]
          f_W ( x \lvert \sigma_1,\nu_2)
        dx\\
&<&1-  \Phi\left(\frac{c_0 -c(t,\nu_2)  }{\sigma_1}
        \right),
\ese
$
\{c_0 -c(t,\nu_2) \}/\sigma_1\le c_1.
$
Thus, $c(t,\nu_2)/\sigma_1\ge
c_0/\sigma_1-c_1\to\infty$. 
On the other hand,
\bse
\alpha_0&=&
\int_0^{c(t,\nu_2)/t} 
\left[\Phi\left(\frac{
c_0+  c(t,\nu_2) - x t}{\sigma_1} \right) 
        -  \Phi\left(\frac{c_0 -c(t,\nu_2) + x  t }{\sigma_1} \right)
          \right]
          f_W ( x\lvert \sigma_1,\nu_2)
        dx\\
&>&\int_0^{c(t,\nu_2)/t} 
\left[\Phi\left(\frac{
c_0}{\sigma_1} \right) 
        -  \Phi\left(\frac{c_0 -c(t,\nu_2) + x  t }{\sigma_1} \right)
          \right]
          f_W ( x \lvert \sigma_1,\nu_2)
        dx\\
&=&\Phi\left(\frac{
c_0}{\sigma_1} \right) 
\int_0^{c(t,\nu_2)/t} 
          f_W ( x \lvert \sigma_1,\nu_2)
        dx-\int_0^{c(t,\nu_2)/t} 
\Phi\left(\frac{c_0 -c(t,\nu_2) + x  t }{\sigma_1} \right)
          f_W ( x \lvert \sigma_1,\nu_2)
        dx\\
&=&\Phi\left(\frac{
c_0}{\sigma_1} \right) 
\int_0^{\{c(t,\nu_2)/(t\sigma_1)\}^2} 
          g(y,\nu_2)
        dy-\int_0^{\{c(t,\nu_2)/(t\sigma_1)\}^2} 
\Phi\left(\frac{c_0 -c(t,\nu_2) }{\sigma_1} +\sqrt{y}t\right)
          g(y,\nu_2)
        dy,
\ese
where
\bse
g(y,\nu_2)
=\frac{\left(\nu_2y
    \right)^{\nu_2/2-1}
  e^{-\nu_2y /
    2}}{2^{\nu_2 / 2} \Gamma\left(\frac{\nu_2}{2}\right)}\nu_2.
\ese
Note that $g(y,\nu_2)$ has mean 1 and variance $2/\nu_2$, hence 
\bse
&&\int_0^{\{c(t,\nu_2)/(t\sigma_1)\}^2} 
\Phi\left(\frac{c_0 -c(t,\nu_2) }{\sigma_1} +\sqrt{y}t\right)
          g(y,\nu_2)
        dy\\
&>&\Phi\left(\frac{
c_0}{\sigma_1} \right) 
\int_0^{\{c(t,\nu_2)/(t\sigma_1)\}^2} 
          g(y,\nu_2)
        dy-\alpha_0\\
&\to&1-\alpha_0
\ese
when $\nu_1, \nu_2\to\infty$, hence for $\nu_1, \nu_2$ sufficiently
large, 
\bse
&&\int_0^{1+\nu_2^{-1/3}}
\Phi\left(\frac{c_0 -c(t,\nu_2) }{\sigma_1} +\sqrt{y}t\right)
          g(y,\nu_2)
        dy \\ 
        &&+
\int_{1+\nu_2^{-1/3}}^{\{c(t,\nu_2)/(t\sigma_1)\}^2} 
\Phi\left(\frac{c_0 -c(t,\nu_2) }{\sigma_1} +\sqrt{y}t\right)
          g(y,\nu_2)
        dy > 1-1.5\alpha_0.
\ese
Now
\bse
&&\int_{1+\nu_2^{-1/3}}^{\{c(t,\nu_2)/(t\sigma_1)\}^2} 
\Phi\left(\frac{c_0 -c(t,\nu_2) }{\sigma_1} +\sqrt{y}t\right)
          g(y,\nu_2)
        dy\\
&\le&\int_{1+\nu_2^{-1/3}}^{\{c(t,\nu_2)/(t\sigma_1)\}^2} 
          g(y,\nu_2)
        dy\\
&\to&0
\ese
when $\nu_2\to\infty$, hence
\bse
\Phi\left(\frac{c_0 -c(t,\nu_2) }{\sigma_1} +(1+\nu_2^{-1/3})t\right)\ge
\int_0^{1+\nu_2^{-1/3}}
\Phi\left(\frac{c_0 -c(t,\nu_2) }{\sigma_1} +\sqrt{y}t\right)
          g(y,\nu_2)
        dy>1-2\alpha_0
\ese
for sufficiently large $\nu_1, \nu_2$. This implies
$
\{c_0-c(t,\nu_2)\}/\sigma_1\ge c_2-(1+\nu_2^{-1/3})t.
$
We thus obtain 
\be\label{eq:bound2}
c_2-  (1+\nu_2^{-1/3})t\le\{c_0-c(t,\nu_2)\}/\sigma_1\le c_1,
\ee
 hence
$|c_0-c(t,\nu_2)|=O(\sigma_1)$.
This leads to $c(t,\nu_2)=c_0+O(\sigma_1)=c_0+o(1)$ and hence
$c(t,\nu_2)/t\asymp1$. 

From a Taylor expansion, we get
\bse
\alpha_0&=&\int_0^{\wh c(t,\nu_2)/t} 
\left[\Phi\left(\frac{
c_0+  \wh c(t,\nu_2) - x t}{\wh\sigma_1} \right) 
        + \Phi\left(\frac{\wh c(t,\nu_2) -c_0-x t }{\wh\sigma_1} \right)-1
          \right]
          f_W ( x\lvert \sigma_1,\nu_2)
        dx\\
&=&\int_0^{c(t,\nu_2)/t} 
\left[\Phi\left(\frac{
c_0+  c(t,\nu_2) - x t}{\sigma_1} \right) 
        +\Phi\left(\frac{c(t,\nu_2)-c_0 - x  t }{\sigma_1} \right)-1
          \right]
          f_W ( x\lvert \sigma_1,\nu_2)
        dx\\
&&+\frac{\wh c(t,\nu_2)-c(t,\nu_2)}{\sigma_1}\int_0^{c(t,\nu_2)/t} 
\left[\phi\left(\frac{
c_0+  c(t,\nu_2) - x t}{\sigma_1} \right) \right. \\
        &&+ \left. \phi\left(\frac{c(t,\nu_2)-c_0 - x  t }{\sigma_1} \right)
          \right]
          f_W ( x\lvert \sigma_1,\nu_2)
        dx\\
&&-\frac{\wh\sigma_1-\sigma_1}{\sigma_1}\int_0^{c(t,\nu_2)/t} 
\left[\phi\left(\frac{c_0+  c(t,\nu_2)-xt}{\sigma_1}\right) 
\frac{c_0+  c(t,\nu_2)-xt}{\sigma_1}\right.\\
&&\left.        + \phi\left(\frac{c(t,\nu_2)-c_0-xt}{\sigma_1}\right)
\frac{c(t,\nu_2)-c_0-xt}{\sigma_1}
          \right]     f_W ( x\lvert \sigma_1,\nu_2)
        dx +r_1\\
&=&\alpha_0+\frac{\wh c(t,\nu_2)-c(t,\nu_2)}{\sigma_1}\int_0^{c(t,\nu_2)/t} 
        \phi\left(\frac{c(t,\nu_2)-c_0 - x  t }{\sigma_1} \right)
          f_W ( x\lvert \sigma_1,\nu_2)
        dx\\
&&-\frac{\wh\sigma_1-\sigma_1}{\sigma_1}\int_0^{c(t,\nu_2)/t} 
\phi\left(\frac{c(t,\nu_2)-c_0-xt}{\sigma_1}\right)
\frac{c(t,\nu_2)-c_0-xt}{\sigma_1}
            f_W ( x\lvert \sigma_1,\nu_2)
        dx +r_2.
\ese
Here we note that both integrals above are bounded, and $r_1, r_2$ are
higher order terms that are ignorable.
Thus, 
\bse
&&|\wh c(t,\nu_2)-c(t,\nu_2)|\\
&\le&2\sigma_1\frac{\frac{|\wh\sigma_1-\sigma_1|}{\sigma_1}
\big|\int_0^{c(t,\nu_2)/t}  
\phi\left(\frac{c_0-c(t,\nu_2)+xt}{\sigma_1}\right)
\frac{c_0-c(t,\nu_2)+xt}{\sigma_1}
            f_W ( x\lvert \sigma_1,\nu_2)
        dx \big|}{\int_0^{c(t,\nu_2)/t} 
        \phi\left(\frac{c(t,\nu_2)-c_0 - x  t }{\sigma_1} \right)
          f_W ( x\lvert \sigma_1,\nu_2)
        dx}\\
&=&O_p(\sigma_1\nu_2^{-1/2})
\frac{
\big|\int_0^{\{c(t,\nu_2)/(t\sigma_1)\}^2} 
\phi\left(\frac{c_0-c(t,\nu_2)}{\sigma_1}+\sqrt{y}t\right)
\left(\frac{c_0-c(t,\nu_2)}{\sigma_1}+\sqrt{y}t\right)
            g(y,\nu_2)
        dy \big|}{\int_0^{\{c(t,\nu_2)/(t\sigma_1)\}^2} 
        \phi\left(\frac{c_0-c(t,\nu_2) }{\sigma_1} +\sqrt{y}t\right)
          g(y,\nu_2)
        dy}.
\ese
Note that
\bse
&&\lim_{\nu_1,\nu_2\to\infty}\frac{
\big|\int_0^{\{c(t,\nu_2)/(t\sigma_1)\}^2} 
\phi\left(\frac{c_0-c(t,\nu_2)}{\sigma_1}+\sqrt{y}t\right)
\left(\frac{c_0-c(t,\nu_2)}{\sigma_1}+\sqrt{y}t\right)
            g(y,\nu_2)
        dy \big|}{\int_0^{\{c(t,\nu_2)/(t\sigma_1)\}^2} 
        \phi\left(\frac{c_0-c(t,\nu_2) }{\sigma_1} +\sqrt{y}t\right)
          g(y,\nu_2)
        dy}\\
&=&\lim_{\nu_1,\nu_2\to\infty}\frac{
\big|
\phi\left(\frac{c_0-c(t,\nu_2)}{\sigma_1}+t\right)
\left(\frac{c_0-c(t,\nu_2)}{\sigma_1}+t\right)
        \big|}{
        \phi\left(\frac{c_0-c(t,\nu_2) }{\sigma_1} +t\right)
          }\\
&=&\lim_{\nu_1,\nu_2\to\infty}\big|\frac{c_0-c(t,\nu_2)}{\sigma_1}+t\big|\\
&=&O(1).
\ese
Thus, 
\bse
\wh c(t,\nu_2)-c(t,\nu_2)=O_p(\sigma_1\nu_2^{-1/2}).
\ese

We also have  $c(t,\nu_2)-t\wh\sigma_1
=c(t,\nu_2) -t\sigma_1+O_p(\sigma_1/\sqrt{\nu_2})
=c_0 -t\sigma_1+O_p(\sigma_1+\sigma_1/\sqrt{\nu_2})
=c_0+o_p(1)$. Thus we indeed have
\bse
\wh c(t,\nu_2)-c(t,\nu_2)=o_p\{c(t,\nu_2)-t\wh\sigma_1\}.
\ese
\end{proof}

\begin{proof}[Proof of result~3 from Proposition~\ref{pro:asymp}]

The rejection region at $t=0$ is
$C\{0, c(0)\} = \{\wh\theta: -c(0)<\wh\theta< c(0)\}$, while when $t\ne0$,
the rejection region is
$ C\{t, c(t)\} =\{\wh\theta: 
t\wh\sigma_1-c(t,\nu_2)
<\wh\theta< c(t,\nu_2)-t\wh\sigma_1\}$. 
Note that the approximation errors of the two approximate rejections
are both
 $\wh c(0)-c(0)=O_p(\sigma_1\nu_2^{-1/2})$ and 
$\wh c(t,\nu_2)-c(t,\nu_2)=O_p(\sigma_1\nu_2^{-1/2})$.
Now we have obtained \eqref{eq:bound1} and \eqref{eq:bound2}, combing
them leads to
\bse
c_1\sigma_1
&\le&(c_1+c_2-2\nu_2^{-1/3}t)\sigma_1\\
&\le&
\{c_1+c_2-(1+\nu_2^{-1/3})t\}\sigma_1+t\sigma_1\{1+O_p(\nu^{-1/2})\}\\
&\le&
c(0)-c(t,\nu_2)+t\wh\sigma_1\\
&\le&(c_1+c_2)\sigma_1+t\sigma_1\{1+O_p(\nu_2^{-1/2})\}\\
&\le&2(c_1+c_2)\sigma_1,
\ese
where we used \eqref{eq:bound1} and \eqref{eq:bound2} to obtain the
third and fourth inequalities above.
This implies the two rejection regions differ by a constant times
$\sigma_1$, i.e., $|c(0)-c(t,\nu_2)+t\wh\sigma_1|\asymp \sigma_1$.
\end{proof}

\section{Proof of Theorem~\ref{th:asy}}
\label{appendix:th:asy}

\begin{proof}[Proof of result~1 from Theorem~\ref{th:asy}]
For any $t \geq 0$, one can easily verify that there exists a positive constant $u$ such that
\bse 
\left\lvert \frac{d}{dc} \omega \{ c_0, \sigma_1, \nu_2,
      t, c\} \right\rvert \leq u .
\ese
Therefore, by the mean value theorem, for any $t \geq 0$, there exists $\wt c (t, \nu_2) \in \left(\wh c(t, \nu_2), c(t, \nu_2) \right)$ such that
\bse
\omega \{ c_0, \sigma_1, \nu_2,
      t, \wh c(t,\nu_2) \}  &= \alpha_0 + \frac{d}{dc} \omega \{ c_0, \sigma_1, \nu_2,
      t, \wt c(t,\nu_2) \} \{\wh c(t,\nu_2)  - c(t,\nu_2) \} ,
\ese
since $\omega \{ c_0, \sigma_1, \nu_2, t, c(t, \nu_2) \}  = \alpha_0$ as $\{ t, c(t)\} \in \calA(\alpha_0)$ in \eqref{eq:adjustments}.
Moreover, we have 
\begin{align*}
\lvert \omega \{ c_0, \sigma_1, \nu_2, t, \wh c(t,\nu_2) \}  - \alpha_0 \rvert &=  
\frac{d}{dc} \omega \{ c_0, \sigma_1, \nu_2, t, \wt c(t,\nu_2) \} \{\wh c(t,\nu_2)  - c(t,\nu_2) \} \\
&\leq u \left \lvert \wh c(t,\nu_2)  - c(t,\nu_2) \right\rvert \\
&=  O_p\left(\sigma_1 / \sqrt{\nu_2} \right) ,
\end{align*}
where the last equality follows from  Proposition~\ref{pro:asymp}.
Therefore, we obtain 
\be \label{eq:order_pf_thm2}
\omega \{ c_0, \sigma_1, \nu_2,
      t, \wh c(t,\nu_2) \} = \alpha_0 + O_p\left(\sigma_1 / \sqrt{\nu_2} \right) . 
\ee
\end{proof}

\begin{proof}[Proof of result~2 from Theorem~\ref{th:asy}]

Using 
$
\omega \{ \theta, \sigma_1, \nu_2,
      0, \wh c(0) \} =  \omega \{ \theta, \sigma_1, \nu_2,
      0, c(0) \}  + O_p (\sigma_1 / \sqrt{\nu_2} ) ,
$
and 
$
\omega \{ \theta, \sigma_1, \nu_2,
      t, \wh c(t,\nu_2) \}  =  \omega \{ \theta, \sigma_1, \nu_2,
      t, c(t,\nu_2) \} + O_p (\sigma_1 / \sqrt{\nu_2} ) ,
$
similarly to \eqref{eq:order_pf_thm2},
we have 
$
\omega \{ \theta, \sigma_1, \nu_2,
      0, \wh c(0) \} - \omega \{ \theta, \sigma_1, \nu_2,
      t, \wh c(t,\nu_2) \}  =  \omega \{ \theta, \sigma_1, \nu_2,
      0, c(0) \} - \omega \{ \theta, \sigma_1, \nu_2,
      t, c(t,\nu_2) \} + O_p\left(\sigma_1 / \sqrt{\nu_2} \right) .
$
As Theorem~\ref{thm:most_powerful} ensures that 
$\omega \{ \theta, \sigma_1, \nu_2, 0, c(0) \} - \omega \{ \theta, \sigma_1, \nu_2, t, c(t,\nu_2) \} \geq 0$, it follows that, for sufficiently small $\sigma_1 / \nu_2$, we have $\omega \{ \theta, \sigma_1, \nu_2, 0, \wh c(0) \} - \omega \{ \theta, \sigma_1, \nu_2, t, \wh c(t,\nu_2) \} \geq 0$.  
\end{proof}

\section{Computational details for univariate cTOST}
\label{appendix:algo_ctost}

Operationally, $\wh c(0)$ can be efficiently computed through the following Newton–Raphson method.
At iteration $k$, with $k \in \mathbb{N}$, we define
\be \label{eq:ctost_iter}
    c^{(k+1)} =  c^{(k)} + \wh\sigma_1\frac{\alpha_0 + \Phi\left(\frac{c_0 - c^{(k)}}{\wh\sigma_1}\right) - \Phi\left(\frac{c_0 + c^{(k)}}{\wh\sigma_1}\right)}{\varphi\left(\frac{c_0 + c^{(k)}}{\wh\sigma_1}\right) + \varphi\left(\frac{c_0 - c^{(k)}}{\wh\sigma_1}\right)},
\ee
where the procedure is initialized at 
\be \label{eq:ctost_iter_init}
\begin{aligned}
    c^{(0)} = I&\left\{c_0 > \wh\sigma_1 \Phi^{-1}(1-\alpha_0)\right\} \left\{c_0 - \wh\sigma_1 \Phi^{-1}(1-\alpha_0)\right\}  \\ 
    &+ I\left\{c_0 \leq \wh\sigma_1 \Phi^{-1}(1-\alpha_0)\right\} c_0.
\end{aligned}
\ee
This choice for $c^{(0)}$ is motivated by the simple approximation $\alpha_0 = \Phi\left\{\frac{c_0+ \wh c(0)}{\wh\sigma_1} \right\} -\Phi\left\{\frac{c_0- \wh c(0)}{\wh\sigma_1} \right\} \approx 1 -\Phi\left\{\frac{c_0- \wh c(0)}{\wh\sigma_1} \right\}$ when $\wh \sigma_1$ is sufficiently small. At convergence of the sequence $c^{(k)}$, we have $c^{(k)} = c^{(k+1)} = \wh c(0)$. 
The procedure to compute the adjustment $\wh c (0)$ is described in
Algorithm~\ref{alg:univ_cTOST}.

\begin{spacing}{1}
\normalem 
\begin{algorithm}[H]
    \SetAlgoLined
      \Input{nominal significance level $\alpha_0$, 
      equivalence margins $c_0$,  
      standard error $\hsigma_1$, degrees of freedom $\nu_2$, algorithmic tolerance $\epsilon_{min}$; }
      \Output{cTOST adjustment $\wh c(0)$; } 
      {Initialize $c^{(0)}$ as in \eqref{eq:ctost_iter_init};} \\ 
      {Obtain $c^{(1)}$ as in \eqref{eq:ctost_iter};} \\
      {Set $k = 0$;} \\
      \While{
      $ \lvert c^{(k+1)} - c^{(k)} \rvert  > \epsilon_{\min}$}{
      {Set $k = k+1$;} \\
      {Obtain $c^{(k+1)}$ as in \eqref{eq:ctost_iter};}
  }             
  Output the cTOST adjustment $\wh c(0) = c^{(k+1)}$\;
\caption{Algorithm for computing the cTOST adjustment $\wh c(0)$.}
 \label{alg:univ_cTOST}
\end{algorithm}
\ULforem 
\end{spacing}
\vskip 1cm

Since the result of a bioequivalence assessment is often visualized through a confidence interval, cTOST allows this interpretation for sufficiently small $\wh \sigma_1$. 
In this case, we have
$c_0 > \wh c(0)$ and we can consider the interval $\wh
\theta \pm \left\{c_0 - \wh c(0)\right\}$, which allows us to use the IIP and accept H$_1$ if the interval falls entirely within the equivalence
margins $(-c_0, c_0)$.

\section{A comparison of multivariate TOST adjustments}
\label{appendix:mvt_example}

\begin{figure}[t!]
    \centering
    \includegraphics[width=1\textwidth]{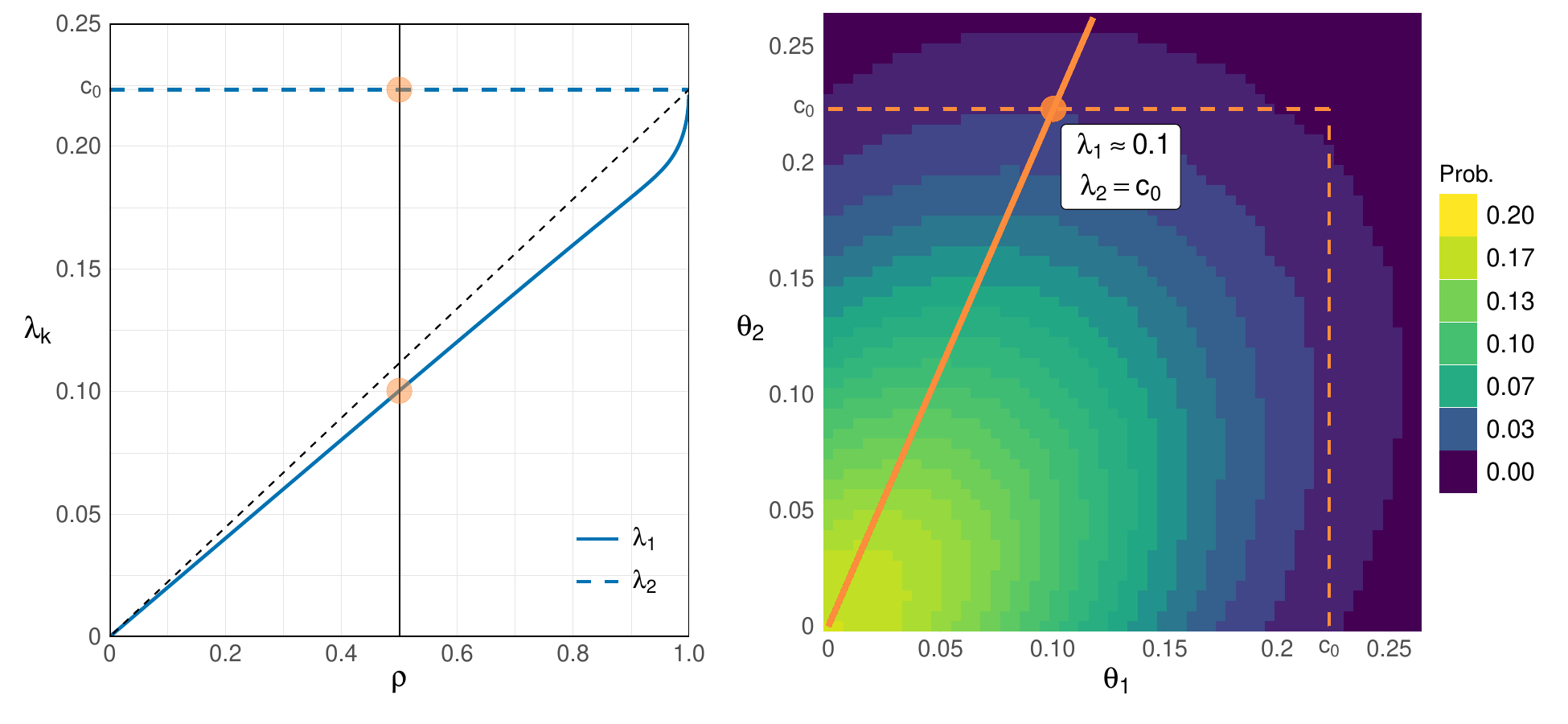}
    \caption{
    Illustrative example of the complex relationship between $\blambda$ in \eqref{eq:lambda_argsup} and the covariance structure of $\hbtheta$ for the bivariate TOST, when $\sigma_{1,1}=\sigma_{1,2}=0.1$, $\nu_2=20$, $c_0=\log(1.25)$, and $\alpha_0=0.05$. 
    Left panel: The $\lambda_1$ coordinate as a function of the correlation level $\rho \in [0,1)$ between the components of $\hbtheta$, while $\lambda_2$ is fixed at $c_0$. Orange dots indicate the coordinates of $\lambda_k$, for $k=1,2$, at $\rho=0.5$. A black dashed line passing through the origin with slope $c_0$ serves as a benchmark.  
    Right panel: Probability that bivariate TOST rejects H$_0$ at $\btheta = (\theta_1,\theta_2)^\top$ in the upper right quadrant when $\rho = 0.5$. The orange dashed lines mark the boundary separating the H$_0$ and H$_1$ regions. The orange dot represents a choice of $\blambda$ (corresponding to $\lambda_2=c_0$, as in the left panel). The orange solid line, connecting the origin to $\blambda$, illustrates the method-specific trajectory of $\btheta$ used for comparing different methods in the numerical study of Section~\ref{sec:simulation_multiv}.} 
    \label{fig:lambda_bivariate}
\end{figure}

First, we illustrate in Figure~\ref{fig:lambda_bivariate} how 
$\blambda(\t,\c)$ is affected by the structure of $\bSigma_1$ when we set $\t=t_{\alpha_0,\nu_2}\1_K, \c=c_0\1_K$ (i.e.,~for the conventional TOST procedure), and consider a simplified setting with $K=2$, $\sigma_{1,1}=\sigma_{1,2}=0.1$, $\nu_2=20$, $c_0=\log(1.25)$, and $\alpha_0=0.05$.

\begin{figure}[ht!]
    \centering    
    \includegraphics[width=0.85\textwidth]{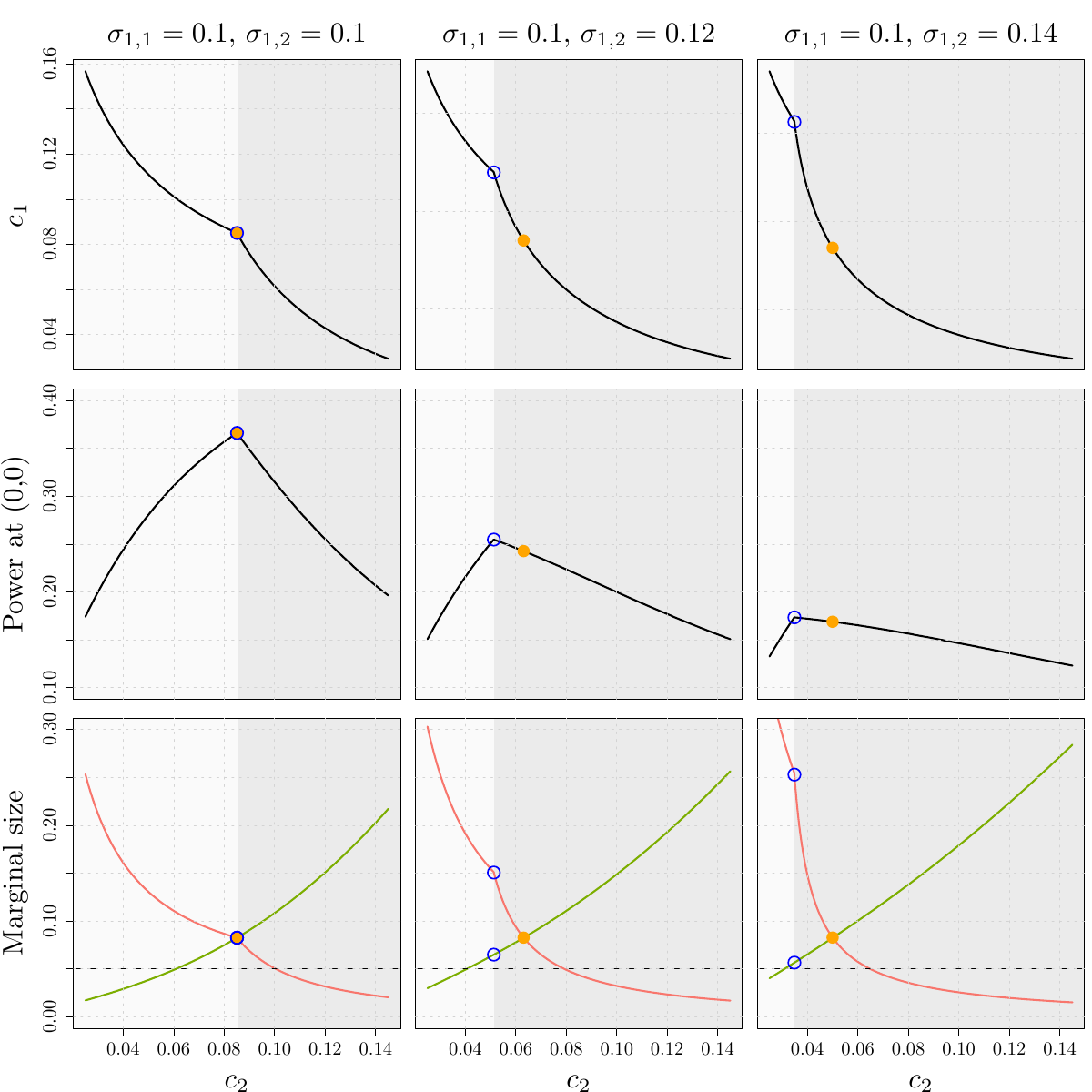}
    \caption{
    Adjustment $c_1(0)$ when considering $\c(\0) = \{ c_1(0), c_2 \}\trans$ such that $\{\0, \c(\0)\} \in \calA_K^*(\alpha_0)$ (first row), corresponding power at $\bt = (0,0)\trans$ (second row), and marginal sizes, 
    respectively highlighted in red and green across the two components,
    where the dashed horizontal line is at the nominal $\alpha_0=5\%$ (third row), 
    across various configurations of
    $\sigma_{1,2} \in \{0.1, 0.12, 0.14\}$ (from left to right columns) when $\sigma_{1,1}=0.1$.
    The multivariate cTOST, 
    $(\0, \c^*)  \in \calA_K(\alpha_0)$, 
    leads to equal marginal sizes, and its operating characteristics are highlighted with orange dots.
    The ``other'' adjustment $\{ \0 , \widetilde\c(\0) \} \in \calA_K^*(\alpha_0)$ maximizes the power at $\bt = (0,0)\trans$, and its operating characteristics are highlighted with blue circles.
    Regions shaded in lighter and darker grays denote the areas where $\blambda$ corresponds to $\blambda_1$ and $\blambda_2$, respectively.}
    \label{fig:2_cs}
\end{figure}

\begin{figure}[t!]
    \centering
    \includegraphics[width=1\textwidth]{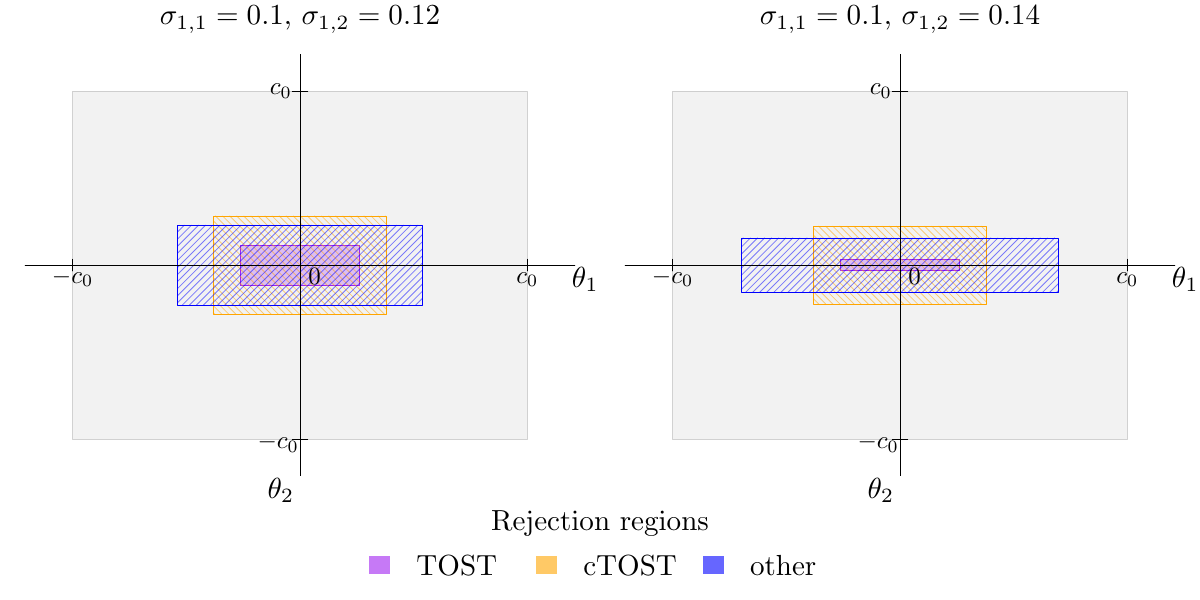}
    \caption{Comparison of the rejection regions of bivariate TOST, cTOST, and the ``other'' potential adjustment based on $\{\0, \widetilde\c(\0)\} \in \calA_K^*(\alpha_0)$,
    when $\sigma_{1,2}=0.12$ (left), $\sigma_{1,2}=0.14$ (right) and
    $\sigma_{1,1}=0.1$.}
    \label{fig:3_rej_reg}
\end{figure}

To compare various methods, for simplicity, we consider a case where $K=2$ and $\bSigma_1$ is a diagonal matrix, with $c_0=\log(1.25)$ and $\alpha_0=0.05$.
We further focus on $\t=\0$ and restrict ourselves to adjustments $\{ \0, \c(\0)\}  \in \calA_K^*(\alpha_0)$.
In this simplified setting, it can be shown that $\blambda$ in \eqref{eq:lambda_argsup} corresponds either to $\blambda_1 \equiv (c_0, 0)\trans $ or $\blambda_2 \equiv (0, c_0)\trans $, 
in the sense that 
$\blambda = \argmax_{\bm{\tau} \in \{\blambda_1, \blambda_2 \} } \omega_K \{ \bm{\tau}, \bSigma_1, \nu_2 , \0,  \c(\0) \} $. 
Moreover, for a fixed value of $c_2$, when a solution to the matching in \eqref{eq:adjustments_mvt} exists, there is a unique adjustment $c_1(0)$ leading to a size of $\alpha_0$.
We thus consider a family of adjustments $\c(\0) = \{ c_1(0), c_2 \}\trans$.
Figure~\ref{fig:2_cs} illustrates how varying $c_2$ affects $c_1(0)$ (first row), the power at $\bt = (0,0)\trans$ (second row), as well as the marginal sizes (third row). 
The columns of Figure~\ref{fig:2_cs} represent different configurations of $\sigma_{1,2} \in \{0.1, 0.12, 0.14\}$ while keeping $\sigma_{1,1}$ fixed at 0.1.
Within each panel, the regions shaded in lighter and darker grays denote the areas where $\blambda$ corresponds to $\blambda_1$ and $\blambda_2$, respectively.
The proposed multivariate cTOST corresponds to the adjustment that leads to equal marginal sizes, that is, 
$(\0, \c^*)  \in \calA_K(\alpha_0)$ in \eqref{eq:adjustments_mvt_constr_size} where $\c^*$ is defined in \eqref{eq:level}-\eqref{eq:level_marginal}, and its operating characteristics are highlighted with orange dots in the panels of Figure~\ref{fig:2_cs}.
We contrast $\c^*$ with another very natural adjustment, say $\{ \0 , \widetilde\c(\0) \} \in \calA_K^*(\alpha_0)$ in \eqref{eq:adjustments_mvt}, corresponding to the one that maximizes the power at $(0,0)$, whose operating characteristics are highlighted with blue circles in the panels of Figure~\ref{fig:2_cs}.
Although the two solutions $\c^*$ and $\widetilde\c(\0)$ coincide under homoskedasticity, they get further apart in the presence of stronger forms of heteroskedasticity (first row).
More specifically, in the latter case, $\c^*$ sacrifices a bit of power (second row) but maintains a balanced marginal risk across all dimensions (third row), in the sense that the type I error rate is evenly split across its components. 
On the other hand, $\widetilde\c(\0)$ may lead to very unbalanced marginal sizes, whose difference is close to 20\% on the third column of Figure~\ref{fig:2_cs}. Thus, the rejection regions associated with $\widetilde\c(\0)$ may be quite uneven across its components.
This is further illustrated in Figure~\ref{fig:3_rej_reg}, where we compare the rejection regions of bivariate TOST, cTOST, and the ``other'' potential adjustment based on $\widetilde\c(\0)$. 
The two panels correspond to the covariance structures respectively used for the second and third column of Figure~\ref{fig:2_cs}.

\section{Equivalence of all adjustments in  $\calA_K(\alpha_0)$ when
\texorpdfstring{$\boldsymbol{\Sigma}_1$}{{\Sigma}_1} is known}
\label{app:known_Sigma_case_mvt}

In this section, we assume that $\bSigma_1$ is known, so that the
(multivariate) canonical form defined in \eqref{eq:mvt_canon} only
relies on the distribution of $\wh\bt$, which is Gaussian.  
Similarly to Appendix~\ref{appendix:power_known_sigma}, we consider a
more flexible counterpart of the multivariate TOST that relies on
$z$-statistics rather than $t$-statistics.
Namely, consider $(\z, \c)$ such that the 
multivariate rejection region of this test is
\be \label{app:mvt_rej_region_known_sigma}
C_K (\z, \c ) = \left\{ \wh\bt \in \real^K
 :\,
\bigcap_{k=1}^K \left( 
z_k \sigma_{1,k}-c_k <\wh\theta_k< c_k-z_k \sigma_{1,k}
\right)\right\},
\ee
where
$\z\equiv(z_1, \dots, z_K)\trans$ 
and $\c\equiv ( c_1, \dots, c_K )\trans$.
The corresponding probability of rejecting H$_0$ is expressed as
\be \label{eq:mvt_proba_known_sigma}
\begin{aligned}
\omega_K\{\bt, \bSigma_1, \infty, \z, \c\}
&=
\Pr\{\wh\bt \in C_K(\z , \c ) \} \\
&=
\Pr \left\{\bigcap_{k=1}^K\left( z_k -\frac{c_k}{\sigma_{1,k}} - \frac{\theta_k}{\sigma_{1,k}} <\frac{\wh\theta_k-\theta_k}{\sigma_{1,k}}< 
\frac{c_k}{\sigma_{1,k}}-z_k - \frac{\theta_k}{\sigma_{1,k}}  \right)\right\} .
\end{aligned}
\ee

Then, 
as a counterpart of \eqref{eq:adjustments_mvt_constr_size}, we restrict ourselves to corrective procedures belonging to the set $\calA_K(\alpha_0)$, such that 
    \be \label{eq:mvt_known_sigma}
    \begin{aligned}
        \calA_K(\alpha_0) = \{(\z, \c):
        \z & \in \real_{\geq0}^K , \c \in \real_{>0}^K, 
        \,
        \omega_K \{ \blambda(\bSigma_1, \z,\c), \bSigma_1 , \infty, \z,  \c\}=
        \alpha_0 ,  \\
         \omega & (c_0, \sigma_{1,1}, \infty, z_1, c_1) = \dots = \omega ( c_0,
        \sigma_{1,K}, \infty,  z_K, c_K )
        \},
    \end{aligned}
    \ee
    and
    \bse 
    \blambda(\bSigma_1, \z,\c) = ( \lambda_1, \ldots, \lambda_K )\trans \in \bm{\Lambda}(\bSigma_1, \z, \c ) \equiv \argsup_{\bt \notin (-c_0, c_0)^K } \;
     \omega_K \{ \bt, \bSigma_1 , \infty, \z,  \c\}.
    \ese

As in the univariate setting described in Appendix~\ref{appendix:power_known_sigma},
    we next show that \eqref{eq:mvt_known_sigma} provides a unique $\c(\z)$
    for any subjectively selected $\z \in \real_{\geq0}^K$.
  Specifically, first, 
    note that $ c_k(\z)$, such that $\{z_k , c_k(\z) \} \in
    \calA(\gamma)$, for $k=1, \ldots, K$, is strictly increasing with
    respect to $\gamma$ (see the discussion following
    \eqref{eq:adjustments}), and hence
    the corresponding probability of rejecting H$_0$, $\omega_K \{
    \bt, \bSigma_1, \nu_2, \z, \c(\z) \}$, is also strictly increasing
    with $\gamma$ at any $\bt \in \real^K$.
    Therefore, 
there exists a unique $\gamma^*$ (see 
    \eqref{eq:appendix:marginal_size_known_sigma} below) such that  
    $
    \omega_K \{ \blambda(\bSigma_1, \z,\c), \bSigma_1 , \infty, \z,  \c(\z)\} = \alpha_0
    $
    in \eqref{eq:mvt_known_sigma}.
Next,
  constraining the marginal sizes to
    \be \label{eq:appendix:marginal_size_known_sigma}
    \omega \{c_0, \sigma_{1,1}, \infty, z_1, c_1(\z)\} = \dots = \omega \{ c_0,
        \sigma_{1,K}, \infty, z_K, c_K(\z) \} = \gamma^*
    \ee
yields a unique rejection region in
    \eqref{app:mvt_rej_region_known_sigma} for any choice of $\z$, due
    to the argument presented in
    Appendix~\ref{appendix:power_known_sigma} for the univariate
    setting. Thus,
    it follows that all corrective procedures $\{\z, \c(\z)\}  \in 
    \calA_K(\alpha_0) $ provide the same rejection region in
    \eqref{app:mvt_rej_region_known_sigma} regardless of the choice of
    $\z$.
    Therefore, these procedures are equivalent in terms of their
    probability of rejecting H$_0$, in the sense that $\omega_K\{\bt,
    \bSigma_1, \infty, \z, \c(\z)\}$, at any $\bt \in \real^K$, does not
    change as $\z$ changes.  

    Finally, the multivariate cTOST in
    \eqref{eq:level}-\eqref{eq:level_marginal} is equivalent to the
    corrective procedures $\{\z, \c(\z)\}  \in \calA_K(\alpha_0)$ in
    \eqref{eq:mvt_known_sigma}. 
    This arises from the fact that its probability of rejecting H$_0$,
    which is $\omega_K(\bt, \bSigma_1, \nu_2, \0, \c^*)$ defined 
    in \eqref{eq:mvt_proba_poplevel}, takes the same form of $
    \omega_K\{\bt, \bSigma_1, \infty, \z, \c(\z)\}$ in
    \eqref{eq:mvt_proba_known_sigma} and,  
    based on the result provided in
    Appendix~\ref{appendix:power_known_sigma}, the marginal rejection
    regions implied by $(0, c_k^*)$ in \eqref{eq:level_marginal} and
    $\{z_k, c_k(\z)\}$ in
    \eqref{eq:appendix:marginal_size_known_sigma} are the same for
    every $k=1,\ldots,K$.  
    Therefore, it follows that
    $\omega_K(\bt, \bSigma_1, \nu_2, \0, \c^*) = \omega_K\{\bt,
    \bSigma_1, \infty,  \0, \c(\0)\} = \omega_K\{\bt, \bSigma_1, \infty, \z,
    \c(\z)\}$  for any $\bt \in \real^K$.

\section{Properties of the multivariate cTOST under independence}
\label{appendix:mvt_ctost}

 In this section, we restrict our attention to adjustments $(\t, \c)
 \in \calA_K(\alpha_0)$ in \eqref{eq:adjustments_mvt_constr_size}. 
We first introduce the following lemma to characterize the set of
parameter values $\bLambda$ in \eqref{eq:lambda_argsup} at which we
evaluate the test size under independence. 
In this case, under both homoskedasticity and heteroskedasticity,
the size is always evaluated at one of the intersections with the coordinate axes.

\begin{lemma} \label{lemma_sup}
Assume that 
$\bSigma_1 = \diag(\sigma_{1,1}^2, \ldots, \sigma_{1,K}^2)$ where
$\diag(\cdot)$ denotes a diagonal matrix, and consider $\{ \t, \c(\t)\} \in
\calA_K(\alpha_0)$ in \eqref{eq:adjustments_mvt_constr_size}. 
It follows that $\bLambda$ in \eqref{eq:lambda_argsup} satisfies
\bse
\bLambda\{\bSigma_1, \nu_2, \t, \c(\t)\}   
    \subseteq
    \{c_0\e_1, \dots, c_0\e_K, -c_0\e_1, \dots, -c_0\e_K\}
    , 
\ese
where 
$\e_h$ is the $h$th standard basis vector in $\real^K$.
\end{lemma}

\begin{proof}[Proof of Lemma~\ref{lemma_sup}]
We recall that $ \{\t, \c(\t)\} \in \calA_K(\alpha_0)$ in
\eqref{eq:adjustments_mvt_constr_size} leads to 
\be
\omega_K \{ \blambda, \bSigma_1, \nu_2, \t, \c(\t) \}
&=
\prod_{k=1}^K
\omega\{\lambda_k, \sigma_{1,k}, \nu_2, t_k, c_k(\t) \} = \alpha_0 
\label{proof_global_size_ct_sup},
\ee
and equal marginal sizes 
\be
\omega\{c_0, \sigma_{1,1}, \nu_2, t_1, c_1(\t) \} = \ldots = \omega\{c_0, \sigma_{1,K}, \nu_2, t_K, c_K(\t) \} 
\label{proof_global_size_ct_marginal} ,
\ee
where the first equality in \eqref{proof_global_size_ct_sup} holds due
to independence.
Let $h$ satisfy
\be \label{eq:lemma_sup_rank}
\omega\{0, \sigma_{1,h}, \nu_2, t_h, c_h(\t) \} \leq 
\omega\{0, \sigma_{1,u}, \nu_2, t_u, c_u(\t) \}
\ee
for all $u\in\{1, \ldots, K\}$, i.e.,~$h$ corresponds to the
smallest marginal power at $\theta=0$.
Note that 
$\omega\{\theta, \sigma_{1,k}, \nu_2, t_k, c_k(\t) \} $ is
non-increasing in $ \theta \in \real_{\geq 0} $, for any $k=1,
\ldots, K$,  
and that to satisfy \eqref{eq:lambda_argsup}
at least one component of $\blambda \equiv \blambda(\t, \c) \in \bLambda\{\bSigma_1, \nu_2, \t, \c(\t)\}$ in \eqref{eq:lambda_argsup} has to be on the boundary $\pm c_0$.
Therefore, since all $K$ components have the same marginal size due to \eqref{proof_global_size_ct_marginal}, 
based on \eqref{proof_global_size_ct_sup} and \eqref{eq:lemma_sup_rank} we get
$$ 
\sup_{\bt \notin (-c_0, c_0)^K } \; \omega_K (\bt, \bSigma_1, \nu_2, \t, \c) =  
\omega\{c_0, \sigma_{1,h}, \nu_2, t_h, c_h(\t) \}
\prod_{\substack{k \in \{1, \ldots, K\},\\k \neq h}}
\omega\{0, \sigma_{1,k}, \nu_2, t_k, c_k(\t) \} .
$$
This implies that the set $\bLambda\{\bSigma_1, \nu_2, \t, \c(\t)\}$
only contains vectors of the form $\blambda_h \equiv c_0 \e_h $, 
where
the $h$th
component satisfies \eqref{eq:lemma_sup_rank}, thus completing the
proof.
\end{proof}

As a direct consequence of Lemma~\ref{lemma_sup}, 
the following corollary provides, as a special case, the set
$\bLambda$ in \eqref{eq:mvt_ctost_sup} at which we evaluate the test size of multivariate cTOST under homoskedasticity.
\begin{cor} \label{cor:sup_ctost}
Assume that 
$\bSigma_1 = \sigma_1^2 \I_K$, and consider $\{ \t, \c(\t)\} \in
\calA_K(\alpha_0)$ in \eqref{eq:adjustments_mvt_constr_size} such that $\t = t \1_K$ with $t \geq 0$. 
Then, $\bLambda$ in \eqref{eq:lambda_argsup} satisfies
\bse 
\bLambda\{\bSigma_1, \nu_2, \t, \c(\t)\}   
    = 
    \{c_0\e_1, \dots, c_0\e_K, -c_0\e_1, \dots, -c_0\e_K\}
    . 
    \ese 
\end{cor}

\begin{proof}[Proof of Corollary~\ref{cor:sup_ctost}]
The result follows immediately from Lemma~\ref{lemma_sup}
since all $K$ components in \eqref{eq:lemma_sup_rank} have same marginal power, in the sense that
$$
\omega\{\theta, \sigma_1, \nu_2, t_1, c_1(\t) \} = \ldots = 
\omega\{\theta, \sigma_1, \nu_2, t_K, c_K(\t) \} ,
$$
for any $\theta \in \real$.
\end{proof}

In the rest of this section, we restrict our attention to the
independent and homoskedastic setting, where $\bSigma_1=\sigma_1^2
\I_K$. 
To assess the statistical power in declaring bioequivalence for adjustments
$\{\t, \c(\t)\} \in \calA_K(\alpha_0)$, that is,  
$\omega_K \{ \bt, \bSigma_1, \nu_2, \t, \c(\t) \}$ defined in \eqref{eq:mvt_proba_poplevel}
for some $\bt \in (-c_0, c_0)^K$,
we begin by establishing Lemma \ref{lemma_mvt}, which 
 will be used in the proofs for
Proposition~\ref{pro:mvt_most_powerful_at_0} and
Proposition~\ref{pro:mvt_no_ump} presented below, where we
respectively consider $\bt = \0$ and $\bt $ in a neighborhood of $ c_0
\1_K$. 

\begin{lemma} \label{lemma_mvt}
Assume that $\bSigma_1 = \sigma_1^2 \mathbf{I}_K$, and consider the
adjustments $\{ \t,
\c(\t)\}$ and $\{\0, \c(\0)\}$ belonging to
$\calA_K(\alpha_0)$ in \eqref{eq:adjustments_mvt_constr_size}. 
The (global) size-$\alpha_0$ constraint in  \eqref{eq:adjustments_mvt_constr_size} respectively leads to 
\begin{align}
\omega_K \{ \blambda_h, \bSigma_1, \nu_2, \t, \c(\t) \}
&=
\omega
\{c_0, \sigma_1, \nu_2, t_h, c_h(\t) \} 
\prod_{\substack{k \in \{1, \ldots, K\},\\k \neq h}}
\omega\{0, \sigma_1, \nu_2, t_k, c_k(\t) \} 
= \alpha_0 ,
\label{proof_global_size_ct}
\\
\omega_K \{ \blambda_g, \bSigma_1, \nu_2, \0, \c(\0) \} &= 
\omega
\{c_0, \sigma_1, \nu_2, 0, c_g(\0)\} 
\prod_{\substack{k \in \{1, \ldots, K\},\\k \neq g}}
\omega
\{0, \sigma_1, \nu_2, 0, c_k(\0) \}
= \alpha_0 ,
\label{proof_global_size_c0}
\end{align}
where $\blambda_k$ (for $k=1, \ldots, K$) is defined in the proof of Lemma~\ref{lemma_sup}, 
$h \in \{1, \ldots, K\}$, and $g = 1, \ldots, K$,
subject to marginal size constraints
\begin{align}
\{t_k, c_k(\t) \} &\in \calA\{\gamma(\t)\} , \label{proof_marg_size_ct} \\
\{0, c_k(\0) \} &\in \calA(\gamma) , \label{proof_marg_size_c0}
\end{align}
for $k=1, \ldots, K$, where $\calA(\cdot)$ is defined in \eqref{eq:adjustments} and $\gamma \equiv \gamma(\0) \le \gamma(\t)$.
\end{lemma}

\begin{proof}[Proof of Lemma~\ref{lemma_mvt}]

  The first equality in
  \eqref{proof_global_size_ct} follows immediately from
  Lemma~\ref{lemma_sup}, which ensures the existence of at least one
  such $h \in \{ 1, \ldots, K \}$ at any $\t$.
The second equality of
\eqref{proof_global_size_ct} follows from the fact that  $\{ \t,
\c(\t) \}$ leads to the global size $\alpha_0$
as it belongs to $\calA_K(\alpha_0)$ in
\eqref{eq:adjustments_mvt_constr_size}. 
Considering the special case of $\t=\0$ leads to 
\eqref{proof_global_size_c0}.

We next show that $\gamma\le\gamma(\t)$ by contradiction.
Assume $\gamma>\gamma(\t)$. 
Note that $\omega
\{c_0, \sigma_1, \nu_2, t_h, c_h(\t) \}=\gamma(\t)$ and
$\omega
\{c_0, \sigma_1, \nu_2, 0, c_g(\0) \}=\gamma$.
For any $k\ne h$, let $\wt c_k(s)$ be the function of $s$ such that
$\omega\{c_0, \sigma_1, \nu_2, s, \wt c_k(s) \}=\gamma(\t)$ for all $s$.
Because $\omega\{c_0, \sigma_1, \nu_2, 0, \wt c_k(0) \}$ is an increasing
function of  $\wt c_k(0)$ (see the discussion following
\eqref{eq:adjustments}),
from
\bse
\gamma=
\omega\{c_0, \sigma_1, \nu_2, 0, c_k(\0) \}>\omega
\{c_0, \sigma_1, \nu_2, 0, \wt c_k(0) \}=\gamma(\t)
=\omega
\{c_0, \sigma_1, \nu_2, t_k, c_k(\t) \},
\ese
we obtain $c_k(\0)> \wt c_k(0)$.
Then
\bse
\omega\{0, \sigma_1, \nu_2, 0, c_k(\0) \}
>\omega\{0, \sigma_1, \nu_2, 0, \wt c_k(0) \}
\ge
\omega\{0, \sigma_1, \nu_2, t_k, c_k(\t) \} ,
\ese
where the first inequality is because of $c_k(\0)>\wt c_k(0)$ and
$\omega\{0, \sigma_1, \nu_2, 0, c_k(\0) \}$ is an increasing function of
$c_k(\0)$,
the second inequality is
due to Theorem \ref{thm:most_powerful}.
Thus, \eqref {proof_global_size_c0} leads to
\bse
\alpha_0&=&\gamma \prod_{\substack{k \in \{1, \ldots, K\},\\k \neq g}}
\omega\{0, \sigma_1, \nu_2, 0, c_k(\0) \}\\
&=&\frac{\gamma \omega\{0, \sigma_1, \nu_2, 0, c_h (\0) \}}{
  \omega\{0, \sigma_1, \nu_2, 0, c_g (\0) \}}
\prod_{\substack{k \in \{1, \ldots, K\},\\k \neq h}}
\omega\{0, \sigma_1, \nu_2, 0, c_k(\0) \}\\
&=&\gamma 
\prod_{\substack{k \in \{1, \ldots, K\},\\k \neq h}}
\omega\{0, \sigma_1, \nu_2, 0, c_k(\0) \}\\
&>& \gamma 
\prod_{\substack{k \in \{1, \ldots, K\},\\k \neq h}}
\omega\{0, \sigma_1, \nu_2, t_k, c_k(\t) \}\\
&=&\gamma/\gamma(\t)\alpha_0,
\ese
which leads to $\gamma(\t)  > \gamma$, contradicting our
assumption. Hence
$\gamma\le\gamma(\t)$.
\end{proof}

We next present  the proof of
Proposition~\ref{pro:mvt_most_powerful_at_0} in
Section~\ref{sec:proposal_multiv}. 
Here, under homoskedasticity and independence, we aim to show that the
probability of declaring bioequivalence at $\bt = \0$,   
within the class of adjustments $\{\t, \c(\t)\} \in \calA_K(\alpha_0)$
in \eqref{eq:adjustments_mvt_constr_size}, is maximized by
multivariate cTOST.

\begin{proof}[Proof of Proposition~\ref{pro:mvt_most_powerful_at_0}]
For $\{\t, \c(\t) \} \in \calA_K(\alpha_0)$, it follows that
\be
\omega
\{0, \sigma_1, \nu_2, t_h, c_h(\t) \} &\leq& \min_{\substack{k \in \{1, \ldots, K\},\\k \neq h}} [ \omega
\{0, \sigma_1, \nu_2, t_k, c_k(\t) \} ]\n\\
&\le&(\prod_{\substack{k \in \{1, \ldots, K\},\\k \neq h}} [ \omega
\{0, \sigma_1, \nu_2, t_k, c_k(\t) \} ])^{1/(K-1)}\n\\
&=& \left(\frac{\alpha_0}{\gamma(\t)}\right)^{\frac{1}{K-1}} , \label{proof_global_size_ct_imply_bis}
\ee
where the first inequality  uses \eqref{eq:lemma_sup_rank}, 
and the last equality uses 
\eqref{proof_global_size_ct} and \eqref{proof_marg_size_ct}. 
Similarly, due to Corollary~\ref{cor:sup_ctost}, for any $k=1,
\dots,K$, 
\be
\omega
\{0, \sigma_1, \nu_2, 0, c_k(\0) \}
= \left(\frac{\alpha_0}{\gamma}\right)^{\frac{1}{K-1}}, 
\ee
based on 
\eqref{proof_global_size_c0} and \eqref{proof_marg_size_c0}.
We thus obtain
\bse
\omega_K \{ \0, \bSigma_1, \nu_2, \t, \c(\t) \} &=&
 \prod_{k=1}^K
\omega
\{0, \sigma_1, \nu_2, t_k, c_k(\t) \} \\
&\le& \left(\frac{\alpha_0}{\gamma(\t)}\right)^{\frac{1}{K-1}}\frac{\alpha_0}{\gamma(\t)}\\
& \le&
\left(\frac{\alpha_0}{\gamma}\right)^{\frac{K}{K-1}} \\
&=&  \prod_{k=1}^K
\omega
\{0, \sigma_1, \nu_2, 0, c_k(\0) \} \\
&=& 
\omega_K \{ \0, \bSigma_1, \nu_2, \0, \c(\0) \}.
\ese
\end{proof}

Despite the optimality of multivariate cTOST at $\bt=\0$,
shown in Proposition~\ref{pro:mvt_most_powerful_at_0},
Proposition~\ref{pro:mvt_no_ump}
illustrates that a uniformly most powerful test for any $\bt \in
(-c_0, c_0)^K$ is generally not achievable in multivariate
settings, even if we only consider the independence and
homoskedasticity setting. This is
very different from the univariate case. 

\begin{pro} \label{pro:mvt_no_ump}
At any given $\nu_2$, $\sigma_1$, consider
$\bSigma_1 = \sigma_1^2 \I_K$.
The family of adjustments $(\t, \c ) \in \calA_K(\alpha_0)$ defined in
\eqref{eq:adjustments_mvt_constr_size}  
does not contain an optimal member $(\t_u, \c_u )$ that satisfies
$$
\omega_K \{ \bt, \bSigma_1, \nu_2, \t, \c \}
\leq 
\omega_K \{ \bt, \bSigma_1, \nu_2, \t_u, \c_u \} ,
$$
for any $\bt \in (-c_0, c_0)^K$.
Therefore, under independence and homoskedasticity, among all (globally) level-$\alpha_0$ tests with equal
marginal sizes, a uniformly most powerful test for the hypotheses in
\eqref{eqn:equiv_hyp_glob_mvt} does not exist.
\end{pro}

\begin{proof}[Proof of Proposition~\ref{pro:mvt_no_ump}]
We prove this result by contradiction.
Assume there exists  an optimal $\t_u$ so that $\{\t_u, \c_u(\t_u)\} \in
\calA_K(\alpha_0)$ satisfies 
\be \label{eq:h_fun_unif_most_powerful}
h(\bt, \sigma_1^2\I_K, \t_u, \t) \equiv 
\omega_K \{ \bt, \sigma_1^2\I_K, \nu_2, \t_u, \c_u(\t_u) \} -
\omega_K \{ \bt, \sigma_1^2\I_K, \nu_2, \t, \c(\t) \}
\geq 0 , \;\;\;\;
\ee
for all $\{\t, \c(\t)\} \in \calA_K(\alpha_0)$ and any $\bt \in (-c_0,
c_0)^K, \sigma_1$, and $\nu_2$.
Hence,
Proposition~\ref{pro:mvt_most_powerful_at_0} and Lemma~\ref{lemma_mvt}
lead to $\gamma(\t_u) = \gamma$.
However, since not all tests achieve the optimal power at
$\bt=\0$ for all $\sigma_1$,
it is possible to find an instance of $\t$ and $\sigma_1$, say
$\t_l$ and $ \sigma_l$,
so that $\{\t_l, \c(\t_l)\} \in
\calA_K(\alpha_0)$, and $h(\0, \sigma_l^2\I_K,\t_u, \t_l) > 0$.
Moreover, let $t_{l,h}$ satisfy
\bse 
\omega\{0, \sigma_{l}, \nu_2, t_{l,h}, c_h(\t_l) \} \leq 
\omega\{0, \sigma_{l}, \nu_2, t_{l,k}, c_k(\t_l) \} ,
\ese
for all $k=1, \ldots, K$,
where $\{ t_{l,k}, c_k(\t_l)\} \in \calA\{\gamma(\t_l)\}$ and $\gamma(\t_l) \geq \gamma$ by Lemma~\ref{lemma_mvt}. 
Hence, to satisfy $h(\0, \sigma_l^2\I_K,\t_u, \t_l) > 0$, it follows that 
$$
\omega\{0, \sigma_{l}, \nu_2, t_{u,k}, c_k(\t_u) \} = \left[\frac{\alpha_0}{\gamma} \right]^{\frac{1}{K-1}}
> \omega\{0, \sigma_{l}, \nu_2, t_{l,h}, c_h(\t_l) \} ,
$$
where the equality is due to Proposition~\ref{pro:mvt_most_powerful_at_0}.
Then, 
note that $\omega\{0, \sigma_l, \nu_2, \t_{l,h}, c_h(\t_l) \}$ is an increasing
function of  $c_h(\t_l)$ (see the discussion following
\eqref{eq:adjustments}),
and consider 
$\t_l^* \equiv t_{l,h} \1_K$ such that $\{ \t_l^*, \c(\t_l^*) \} \in \calA_K(\alpha_0)$.
This leads to 
$\{ t_{l,k}^*, c_k(\t_l^*)\} \in \calA\{\gamma(\t_l^*)\}$, for all $k=1, \ldots, K$, with $\gamma(\t_l^*) > \gamma$.
Therefore, we obtain 
$$
h(\0, \sigma_l^2\I_K,\t_u, \t_l^*) = \left[\frac{\alpha_0}{\gamma} \right]^{\frac{K}{K-1}} - 
\left[\frac{\alpha_0}{\gamma(\t_l^*)} \right]^{\frac{K}{K-1}} > 0 .
$$ 

Now, we have
$h(c_0\1_K, \sigma_l^2 \I_K, \t_u, \t_l^*)
=\gamma^K-\gamma(\t_l^*)^K<0$.
Hence at a point
  $(c_0-\epsilon)\1_K$, where $\epsilon>0$ is a small value, which is very
  close to $c_0\1_K$, we also have
  $h\{(c_0-\epsilon)\1_K, \sigma_l^2  \I_K, \t_u, \t_l^*\}
  =\gamma^K-\gamma(\t_l^*)^K<0$ due to continuity, which is contradictory to the
  optimality of $\t_u$ at any $\btheta\in(-c_0,c_0)^K$ and any $\sigma_1$.
\end{proof}

As a direct consequence of Proposition~\ref{pro:mvt_no_ump}, the
following corollary ensures that a uniformly most powerful test for
any $\bt \in (-c_0, c_0)^K$ is generally not achievable in multivariate
settings.

\begin{cor} \label{cor:mvt_unif}
The family of adjustments $(\t, \c ) \in \calA_K(\alpha_0)$ defined in
\eqref{eq:adjustments_mvt_constr_size}  
does not contain an optimal member $(\t_u, \c_u )$ that satisfies
$$
\omega_K \{ \bt, \bSigma_1, \nu_2, \t, \c \}
\leq 
\omega_K \{ \bt, \bSigma_1, \nu_2, \t_u, \c_u \} ,
$$
for any $\bt \in (-c_0, c_0)^K$ at any $\nu_2, \bSigma_1$.
Therefore, among all (globally) level-$\alpha_0$ tests with equal
marginal sizes, a uniformly most powerful test for the hypotheses in
\eqref{eqn:equiv_hyp_glob_mvt} does not exist.
\end{cor}
\begin{proof}[Proof of Corollary~\ref{cor:mvt_unif}]
The result follows immediately from Proposition~\ref{pro:mvt_no_ump}.
\end{proof}

\begin{figure}[ht]
    \centering
    \includegraphics[width=1\textwidth]{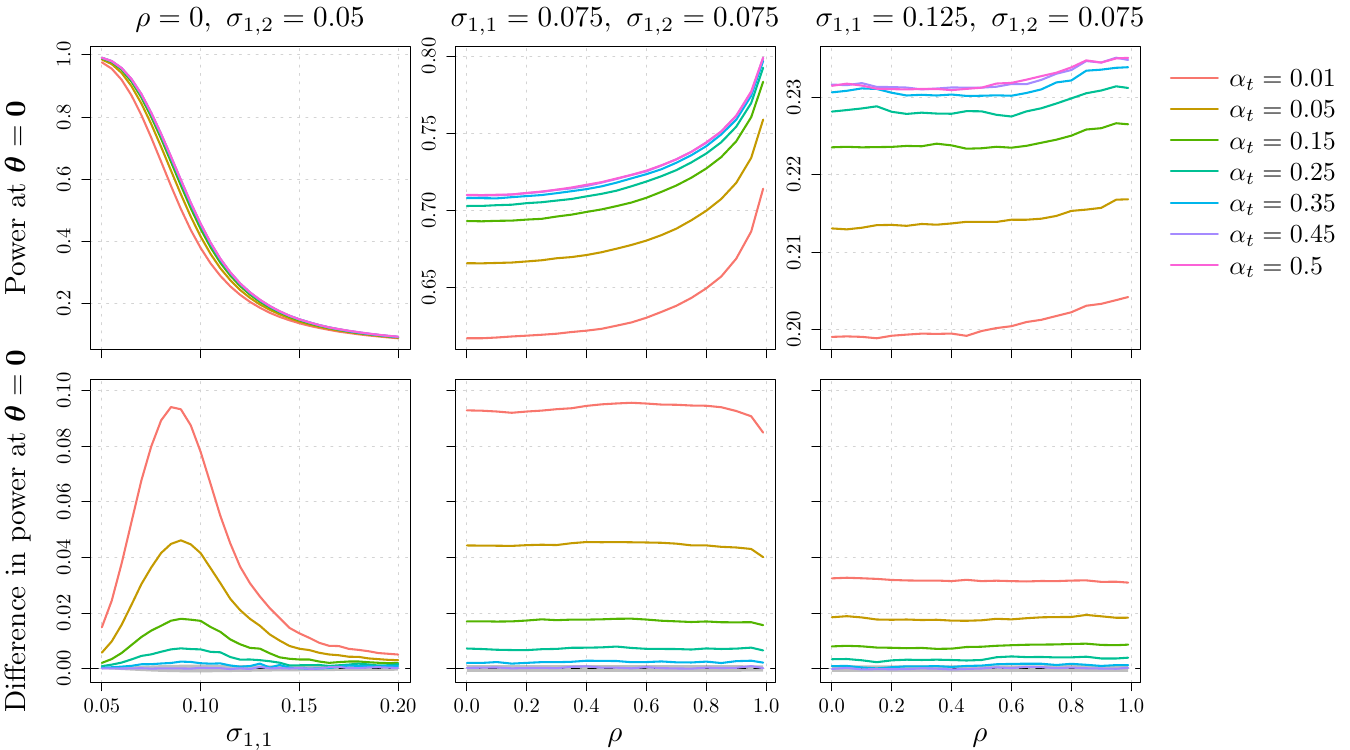}
    \caption{
    Comparison of the power at $\bt=\0$ of various multivariate TOST adjusted procedures (first row), and their difference in power with respect to multivariate cTOST  (second row), under independence and heteroskedasticity (first column), dependence and homoskedasticity (second column), and dependence and heteroskedasticity (third column). 
    }
    \label{fig:conj_fig}
\end{figure}

Although Corollary~\ref{cor:mvt_unif} shows that a uniformly
most powerful test does not exist in
$\calA_K(\alpha_0)$, we conjecture that the optimality of the
multivariate cTOST at $\bt=\0$ 
holds in situations more general than $\sigma_1^2\I_K$,
as required in Proposition~\ref{pro:mvt_most_powerful_at_0}.
This conjecture is supported by extensive numerical results, and in
Figure~\ref{fig:conj_fig} we present the ones related to a bivariate
setting. 
Here we compare adjustments $\{ \t, \c(\t) \} \in \calA_K(\alpha_0)$ for various choices of $\t$.
Specifically, we express $\t = t_{\alpha_t, \nu_2} \1_K $ and vary
$\alpha_t \in \{0.01, 0.05, 0.1, \ldots, 0.45, 0.5 \}$.  
Therefore, the multivariate cTOST $\{\0, \c(\0) \}$ defined in
\eqref{eq:level}-\eqref{eq:level_marginal} is retrieved at $\alpha_t =
0.5$, and at $\alpha_t = \alpha_0$ we obtain a multivariate extension
of the $\delta$-TOST procedure considered in
\citep{boulaguiem2024finite}. 
Specifically, fixing $\alpha_0=0.05$, $\nu_2 = 20$, and
$c_0=\log(1.25)$, in Figure~\ref{fig:conj_fig} we compare the power
$\omega_K \{ \0, \bSigma_1, \nu_2, \t, \c(\t) \}$ of such adjusted
procedures (first row), as well as the difference in power with
respect to multivariate cTOST, that is, 
 $h(\0, \bSigma_1, \0, \t)$ defined in \eqref{eq:h_fun_unif_most_powerful} 
(second row).
We consider the following settings. 
(i) Independent setting under heteroskedasticity (first column), where
$\rho=0$, $\sigma_{1,2}=0.05$ and $\sigma_{1,1}$ increases from 0.05 to 0.2
with a step size of 0.005.  
(ii) Dependent setting under homoskedasticity (second column), where
$\sigma_{1,1} = \sigma_{1,2}=0.075$ and $\rho \in \{0, 0.05, \ldots, 0.9,
0.95, 0.99 \}$. 
(iii) Dependent setting under heteroskedasticity (third column), where
$\sigma_{1,1} = 0.125$, $\sigma_{1,2}=0.075$ and $\rho \in \{0, 0.05, \ldots,
0.9, 0.95, 0.99 \}$.  
In all considered scenarios, multivariate cTOST is always
more powerful than other adjusted methods. 
In all considered scenarios, multivariate cTOST 
appears to provide superior or comparable power at $\0$, in the sense
that, $h(\0, \bSigma_1, \0, \t) \geq 0 $ after accounting for the
simulation error. Indeed, 
differences in power 
(second row of Figure~\ref{fig:conj_fig})
are larger than zero 
for smaller $\alpha_t$'s.
As $\alpha_t$ approaches the optimal value $0.5$, the power gain naturally decreases and the difference  eventually shrinks to within simulation error margins (at a 5\% significance),
represented as gray regions around a power difference of zero,
indicated by black dashed lines.

\cleardoublepage
\newpage
\section{Multivariate cTOST versus multivariate $\alpha$-TOST}
\label{appendix:mvt_ctost_vs_atost}

It is of interest to highlight the advantages of the proposed
multivariate cTOST over the multivariate $\alpha$-TOST \citep{boulaguiem2025mvt}, which can be considered as the current state-of-the-art method. 
The latter arises as a special case of the family of adjustments
$\{\t,\c(\t)\} \in \calA_K^*(\alpha_0)$ in \eqref{eq:adjustments_mvt}
and, therefore, does not
generally lead to equal marginal sizes. In particular, similarly to
the multivariate TOST, it is obtained when $\c = c_0\1_K$, but it replaces $t_{\alpha_0}\1_K$ with
$t_{\alpha^*}\1_K$, where the same adjusted significance level $\alpha^*$
is employed across all $K$ components to ensure uniqueness and that the resulting
adjustment belongs to $\calA_K^*(\alpha_0)$. 
Notably, the operating characteristics of the multivariate
$\alpha$-TOST and the multivariate cTOST are expected to be quite
similar under equi-correlation and homoskedasticity (see Section~\ref{appendix:sim_mvt_main}). 
However, the multivariate cTOST is preferable when these 
assumptions are not satisfied, as the multivariate $\alpha$-TOST
adjusts a scalar $\alpha^*$ across all dimensions and thus lacks flexibility.
Moreover, the cTOST is computationally much leaner than the multivariate $\alpha$-TOST, which requires the computation of additional integrals in \eqref{eq:mvt_proba_poplevel}.

The results of the case study in Section~\ref{sec:application} highlight the higher flexibility of multivariate cTOST, which effectively leverages information across components by applying greater shrinkage to confidence intervals for dimensions associated with larger variance (where
bioequivalence is more difficult to establish) and less shrinkage to dimensions with smaller variance (where it is easier to
establish bioequivalence). 
Overall, ensuring equal marginal size leads to confidence intervals that remain more comparable across all dimensions. 

\section{Computational details for multivariate cTOST}
\label{appendix:algo_ctost_mvt}

The algorithm to construct multivariate cTOST adjustments $\c^*$ in \eqref{eq:level}-\eqref{eq:level_marginal} exploits the fact that for a fixed $\blambda(\c)$ in \eqref{eq:mvt_ctost_sup}, one can easily obtain adjustments $\c(\blambda)$, by recursively applying Algorithm~\ref{alg:univ_cTOST} across each of the $K$ dimensions in $\c$, where the target significance level $\alpha_0$ is replaced by $\gamma \geq \alpha_0$ to ensure that all marginal sizes are equal and at the same time the global size remains controlled at $\alpha_0$.  
Specifically, we match the marginal sizes to a $\gamma$ such that the resulting marginal adjustments $(0, c_k^*) \in \calA(\gamma)$, for $k=1, \ldots, K$, lead to the targeted multivariate size $\omega_K \{ \blambda(\0,\c^*), \bSigma_1, \nu_2 , \0,  \c^* \} = \alpha_0$. For a fixed $\blambda(\0,\c)$, this algorithm can be shown to converge exponentially fast. 
However, as $\blambda(\0,\c)$ generally depends on the adjustments themselves when $K>1$, we consider a recursive optimization scheme. 
Namely, at iteration $r$, with $r \in \mathbb{N}$, we construct the adjustments
\begin{align*}
\c^{(r+1)} = \argzero_{\c} & \; \omega_K\{\blambda(\0,\c^{(r)}), \bSigma_1, \nu_2 , \0,  \c \} - \alpha_0 \\
\text{s.t. } & \omega (c_0, \sigma_{1,1}, \nu_2, 0, c_1) = \dots = \omega ( c_0,
\sigma_{1,K}, \nu_2, 0, c_K ) ,
\end{align*}
where the procedure is initialized at $\c^{(0)} = \c_0$ and iterated until convergence.
The formal procedure to compute $\c^*$ is described in Algorithm~\ref{alg:multiv_cTOST}.

Similarly to the univariate framework, we can expect $c_0 > c_k^*$,
for all $k=1, \ldots, K$, when $\sigma_1$ and $K$ are sufficiently
small. Therefore, we can consider the intervals $\htheta_k \pm \{c_0 -
c_k^* \}$ and, based on the IIP, accept H$_1$ if all $K$ intervals are
entirely contained within the equivalence margins $(-c_0, c_0)$.

\vskip 0.5cm
\begin{spacing}{1}
\normalem 
\begin{algorithm}[H] 
    \SetAlgoLined
      \Input{nominal significance level $\alpha_0$, equivalence margins $c_0$, 
      covariance matrix $\bSigma_1$, degrees of freedom $\nu_2$, maximum number of iterations $r_{\max}$, algorithmic tolerance $\epsilon_{min}$; }
      \Output{multivariate cTOST adjustments $\c^*$; } 
       {Set $r = 0$;} \\
      {Initialize $\c^{(0)} = \c_0$;} \\ 
      {Obtain $\blambda\{ \c^{(0)} \} $ as in \eqref{eq:mvt_ctost_sup};} \\
      \While{
      $ \left\lvert \omega_K [ \blambda \{ \c^{(r)} \}, \bSigma_1, \nu_2 , \0, \c^{(r)}  ] -\alpha_0 \right\rvert > \epsilon_{\min}$ and $r \leq r_{\max}$ }{
      {Initialize $\gamma^{(0)} = \alpha_0$;} \\
      {Obtain $ \c^{(r,0)} $ with Algorithm~\ref{alg:univ_cTOST}
      where $\{ 0, c_k^{(r,0)}\} \in \calA(\gamma^{(0)})$, for $k=1,\ldots,K$;} \\
      {Obtain $ \gamma^{(1)} = \gamma^{(0)} + \alpha_0 - \omega_K [ \blambda \{ \c^{(r)} \}, \bSigma_1, \nu_2 , \0, \c^{(r,0)}  ] $;} \\
      {Initialize $u=0$;} \\
      \While{ $ \lvert \gamma^{(u+1)} - \gamma^{(u)} \rvert > \epsilon_{\min}$ }{
        {Set $u = u+1$;} \\
        {Obtain $ \gamma^{(u+1)} = \gamma^{(u)} + \alpha_0 - \omega_K [ \blambda \{ \c^{(r)} \}, \bSigma_1, \nu_2 , \0, \c^{(r,u)}  ] $;} \\
        {Obtain $ \c^{(r,u+1)} $ with Algorithm~\ref{alg:univ_cTOST},
        where $\{ 0, c_k^{(r,u+1)}\} \in \calA(\gamma^{(u+1)})$, for $k=1,\ldots,K$;} 
      }
      {Set $r = r+1$;} \\
      {Store the adjustment $\c^{(r)} = \c^{(r,u)}$;} \\
      {Compute $\blambda\{ \c^{(r)} \} $ in \eqref{eq:mvt_ctost_sup};}
  }             
  Output the multivariate cTOST adjustment $\c^* = \c^{(r)}$
\caption{Algorithm for computing the multivariate cTOST adjustments $\c^*$}
\label{alg:multiv_cTOST}
\end{algorithm}
\ULforem 
\end{spacing}

\section{Extended univariate simulation results}
\label{appendix:sim_univ_main}

In this section, we present the results of a more extensive simulation study for the univariate setting presented in Section~\ref{sec:simulation_univ}.

\subsection{Setting 1}

We first consider the canonical form defined in \eqref{eq:mvt_canon} for $K=1$, where $34$ equally-spaced values from the interval $[0, 0.26]$ are considered for $\theta$, including the value of $c_0=\log(1.25)$ which provides the value at which we compute the size of any given method as defined in \eqref{eq:tost_size}. 
For the standard error of $\htheta$ we consider $\sigma_1 \in \{ 0.08,0.12,0.16 \} $, and vary the degrees of freedom $\nu_2 \in \{ 5, 10, 15, 20, 30, 40 \}$. All tests are conducted at the nominal significance level $\alpha_0 = 5\%$, based on $10^5$ Monte Carlo samples.

\begin{figure}[ht]
        \centering
        \includegraphics[width=1\textwidth]{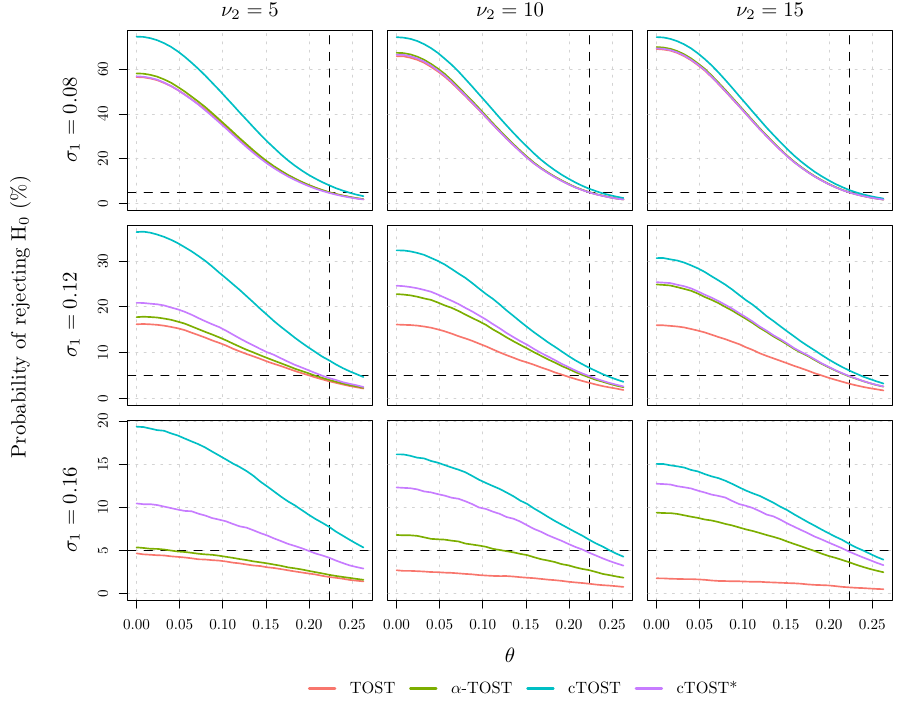}
        \caption{Probability of rejecting the null hypothesis as a function of $\theta$, for the TOST, $\alpha$-TOST, cTOST and cTOST*, across different degrees of freedom $\nu_2$ (columns) and standard errors (rows).}
        \label{fig:power_curves_1}
\end{figure}

\begin{figure}[ht]
        \centering
        \includegraphics[width=1\textwidth]{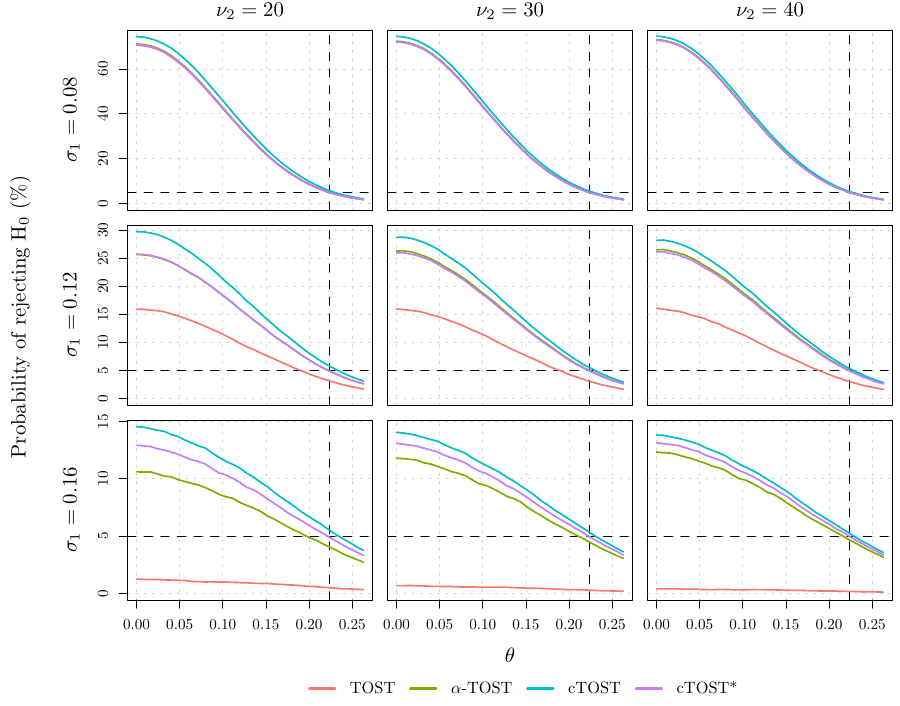}
        \caption{Probability of rejecting the null hypothesis as a function of $\theta$, for the TOST, $\alpha$-TOST, cTOST and cTOST*, across different degrees of freedom $\nu_2$ (columns) and standard errors (rows).}
        \label{fig:power_curves_2}
\end{figure}

Figure~\ref{fig:power_curves_1} compares the probability of rejecting H$_0$ as a function of $\theta$ for the TOST, $\alpha$-TOST, cTOST and cTOST*, where $\nu_2 \in \{ 5, 10, 15\}$.
Here cTOST* consistently achieves an empirical size closest to the nominal level $\alpha_0$ while simultaneously demonstrating uniformly greater power than the TOST and $\alpha$-TOST.
While cTOST is most powerful across all simulation settings, it leads to a slightly liberal test procedure for smaller sample sizes ($\nu_2 < 15$). 
Similar results hold for settings with larger degrees of freedom $\nu_2\in \{ 20, 30, 40\}$ as shown in Figure~\ref{fig:power_curves_2}.

\clearpage
\subsection{Setting 2}

\begin{figure}[ht]
    \centering
    \includegraphics[width=1\textwidth]
    {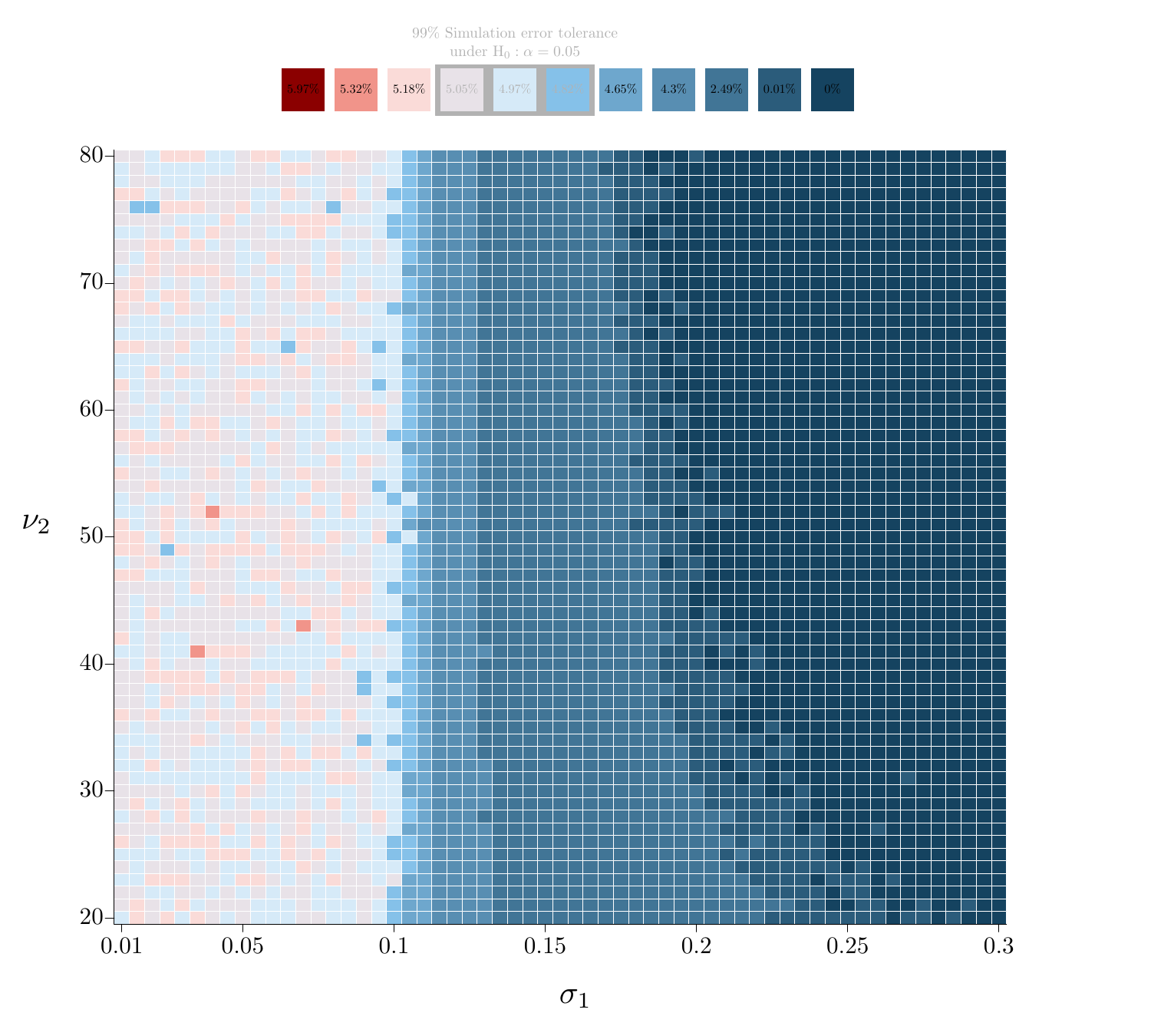}
    \caption{Heatmap representing the empirical size (in \%) of the TOST (color gradient) as a function of $\sigma_1$ ($x$-axis) and $\nu_2$ ($y$-axis). The lighter colors, highlighted in gray in the legend, correspond to the $\alpha_0= 5\%$ nominal level, up to a simulation error. This error was assessed by a two-sided binomial exact test at the 1\% level, based on results from $10^5$ Monte Carlo samples per setting.}
    \label{fig:size_tost_tiles}
\end{figure}

\begin{figure}[ht]
    \centering
    \includegraphics[width=1\textwidth]
    {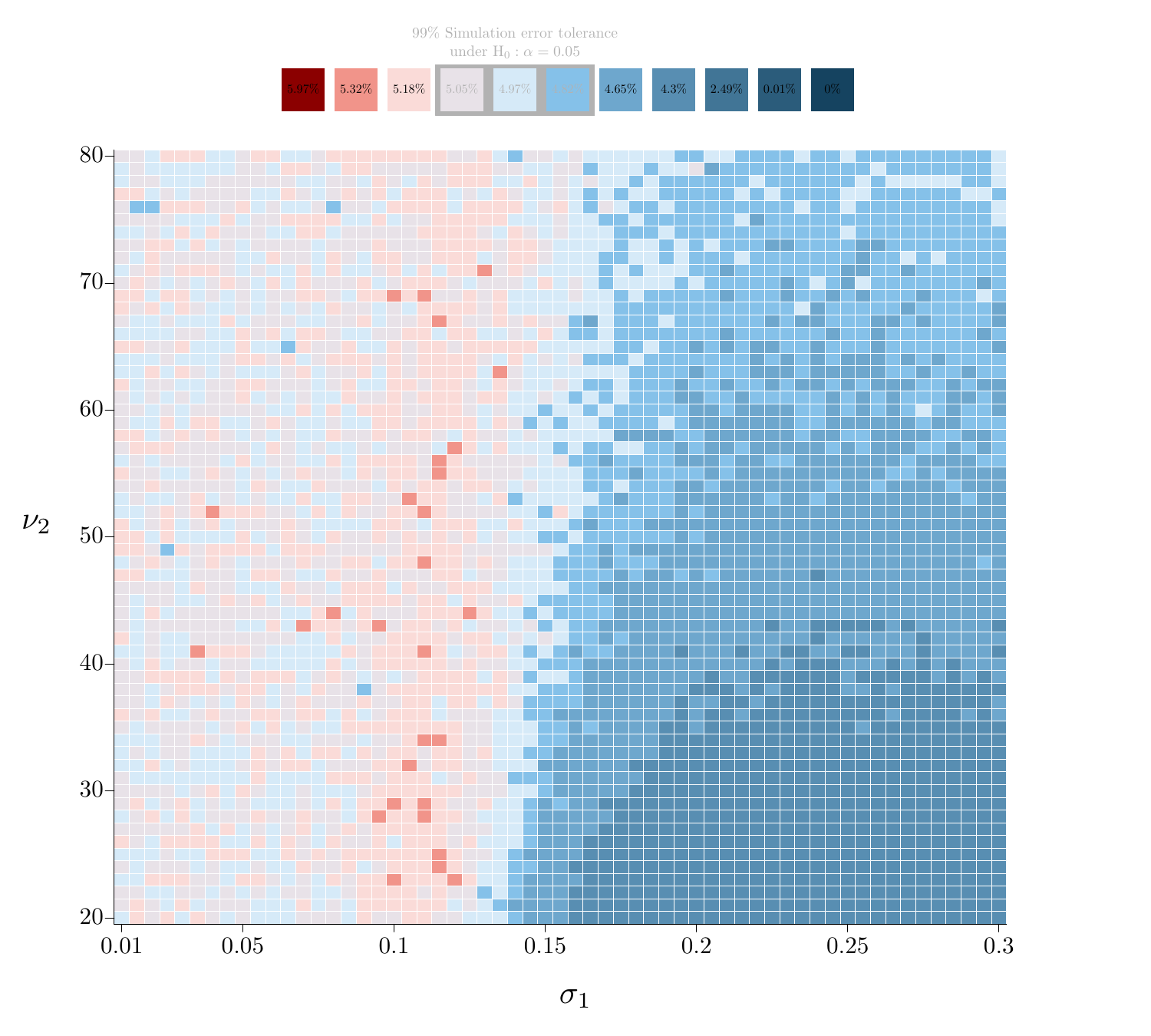}
    \caption{Heatmap representing the empirical size (in \%) of the $\alpha$-TOST (color gradient) as a function of $\sigma_1$ ($x$-axis) and $\nu_2$ ($y$-axis). The lighter colors, highlighted in gray in the legend, correspond to the $\alpha_0= 5\%$ nominal level, up to a simulation error. This error was assessed by a two-sided binomial exact test at the 1\% level, based on results from $10^5$ Monte Carlo samples per setting.}
    \label{fig:size_atost_tiles}
\end{figure}

\begin{figure}[ht]
    \centering
    \includegraphics[width=1\textwidth]
    {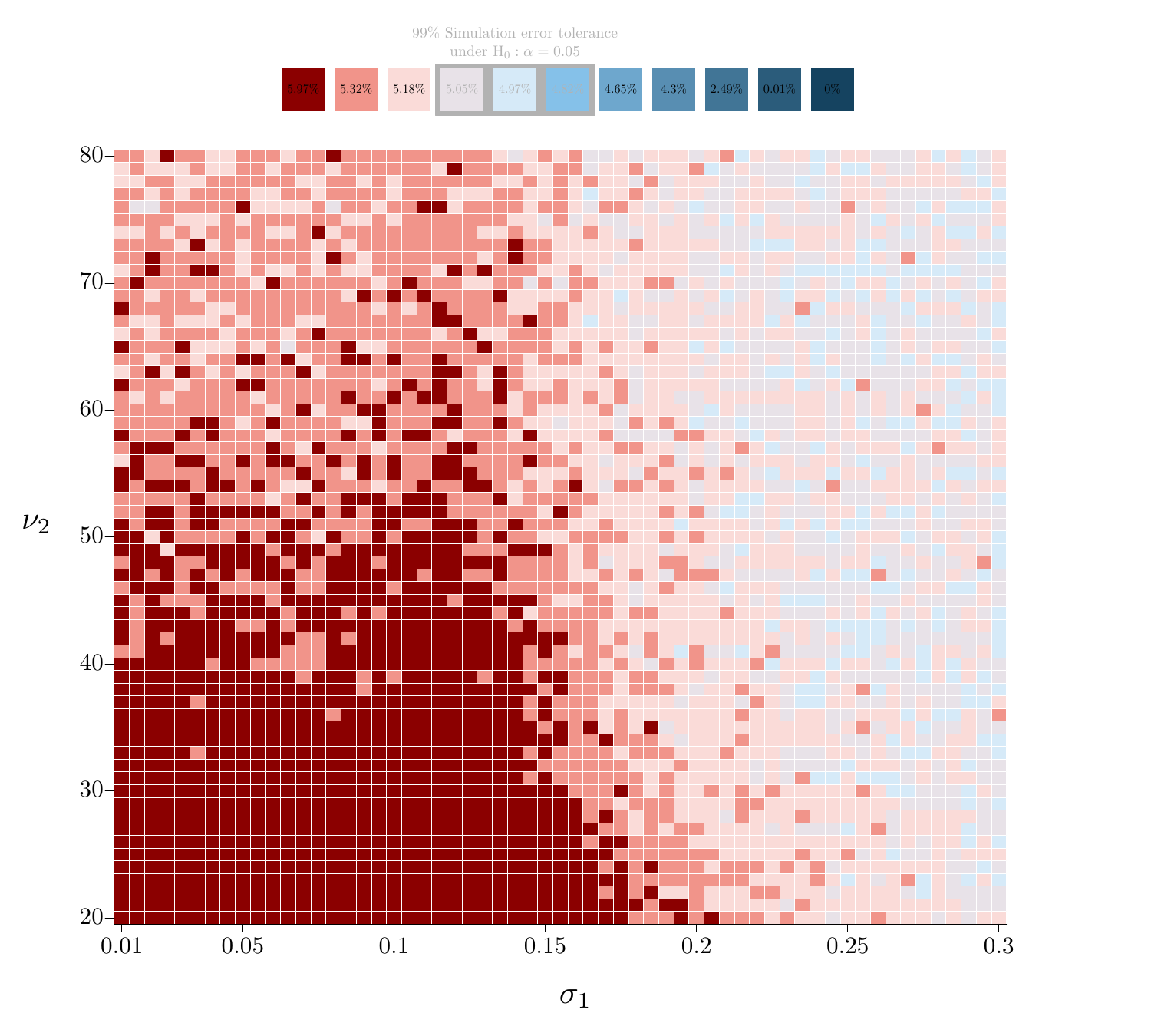}
    \caption{Heatmap representing the empirical size (in \%) of the cTOST (color gradient) as a function of $\sigma_1$ ($x$-axis) and $\nu_2$ ($y$-axis). The lighter colors, highlighted in gray in the legend, correspond to the $\alpha_0= 5\%$ nominal level, up to a simulation error. This error was assessed by a two-sided binomial exact test at the 1\% level, based on results from $10^5$ Monte Carlo samples per setting.}
    \label{fig:size_ctost_tiles}
\end{figure}

\begin{figure}[ht]
    \centering
    \includegraphics[width=1\textwidth]
    {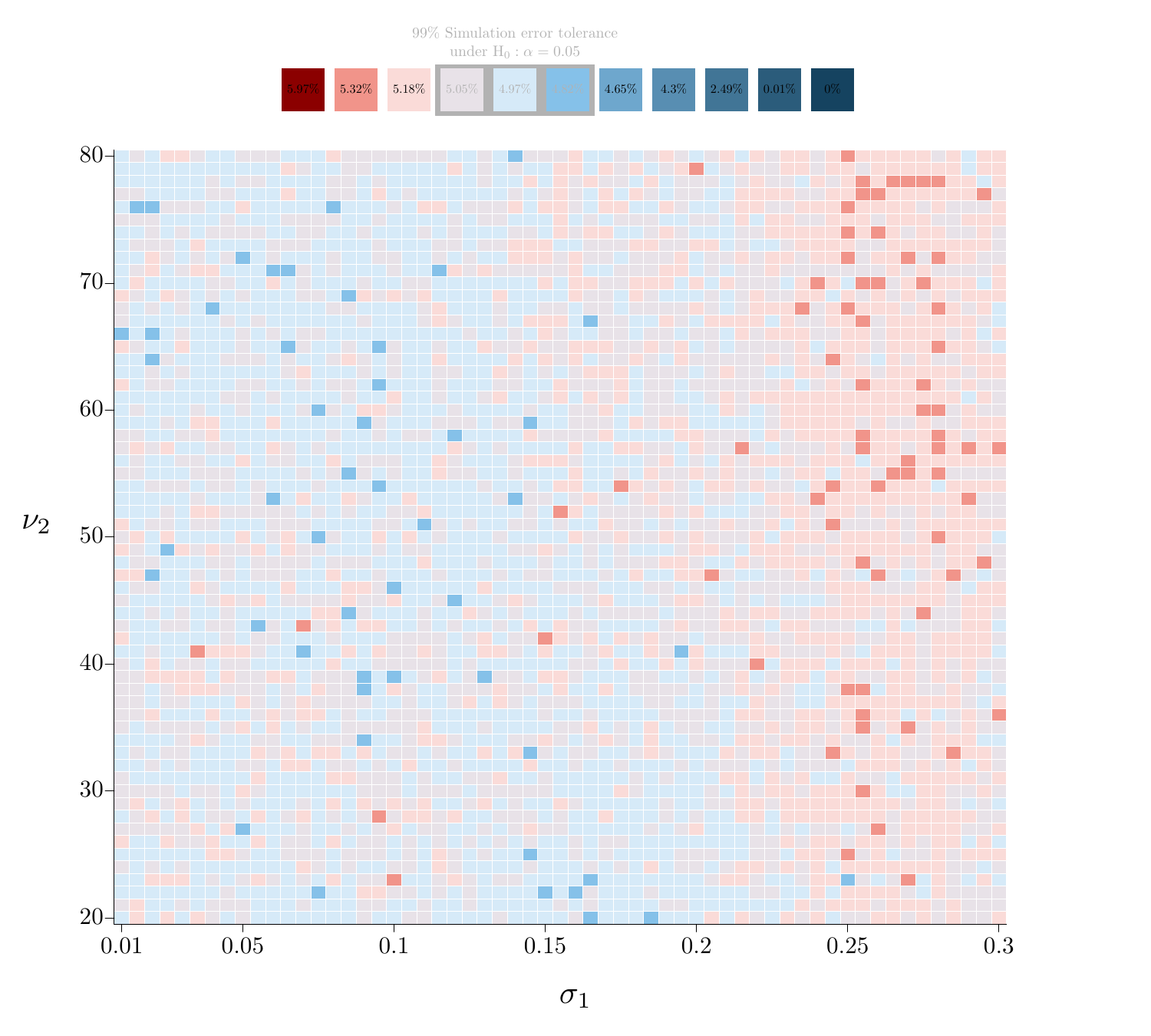}
    \caption{Heatmap representing the empirical size (in \%) of the cTOST* (color gradient) as a function of $\sigma_1$ ($x$-axis) and $\nu_2$ ($y$-axis). The lighter colors, highlighted in gray in the legend, correspond to the $\alpha_0= 5\%$ nominal level, up to a simulation error. This error was assessed by a two-sided binomial exact test at the 1\% level, based on results from $10^5$ Monte Carlo samples per setting.}
    \label{fig:size_ctost_ref_tiles}
\end{figure}

Next, we compare the empirical size of various test procedures across a variety of settings. 
We set $c_0=\log(1.25)$, $\alpha_0=5\%$, and increase $\sigma_1$ from 0.01 to 0.3 with a step size of 0.005, and $\nu_2$ from 20 to 80 with a step size of 1. To evaluate the empirical test size, we fix $\theta = c_0$.
Figures~\ref{fig:size_tost_tiles}-\ref{fig:size_ctost_ref_tiles} show simulation results respectively for TOST, $\alpha$-TOST, cTOST and cTOST*.
Across all simulation scenarios, while
cTOST has a slightly liberal size for smaller $\nu_2$'s and $\sigma_1$'s,
cTOST* consistently maintains a size closest to the nominal level $\alpha_0$. In contrast, both TOST and $\alpha$-TOST procedures often lead to conservative tests.

\begin{figure}
        \centering
        \includegraphics[width=1\textwidth]{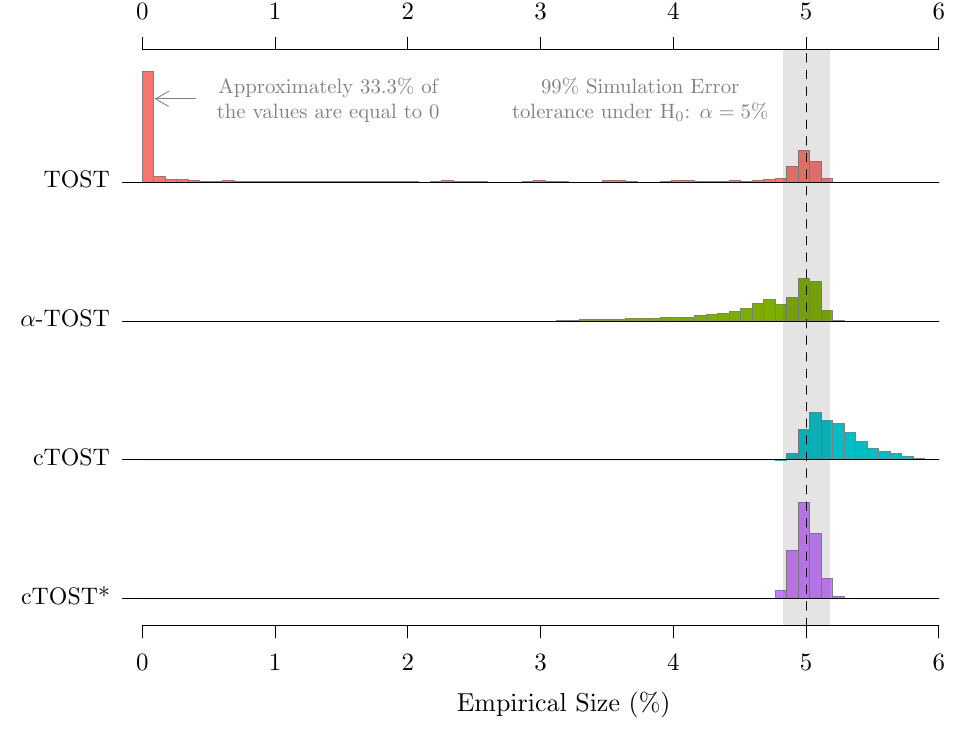}
        \caption{Histograms reporting the empirical test size of TOST, $\alpha$-TOST, cTOST and cTOST* when varying $\sigma_1$ and $\nu_2$.}
        \label{fig:hist_tier}
\end{figure}

\begin{figure}[ht]
        \centering
        \includegraphics[width=1\textwidth]{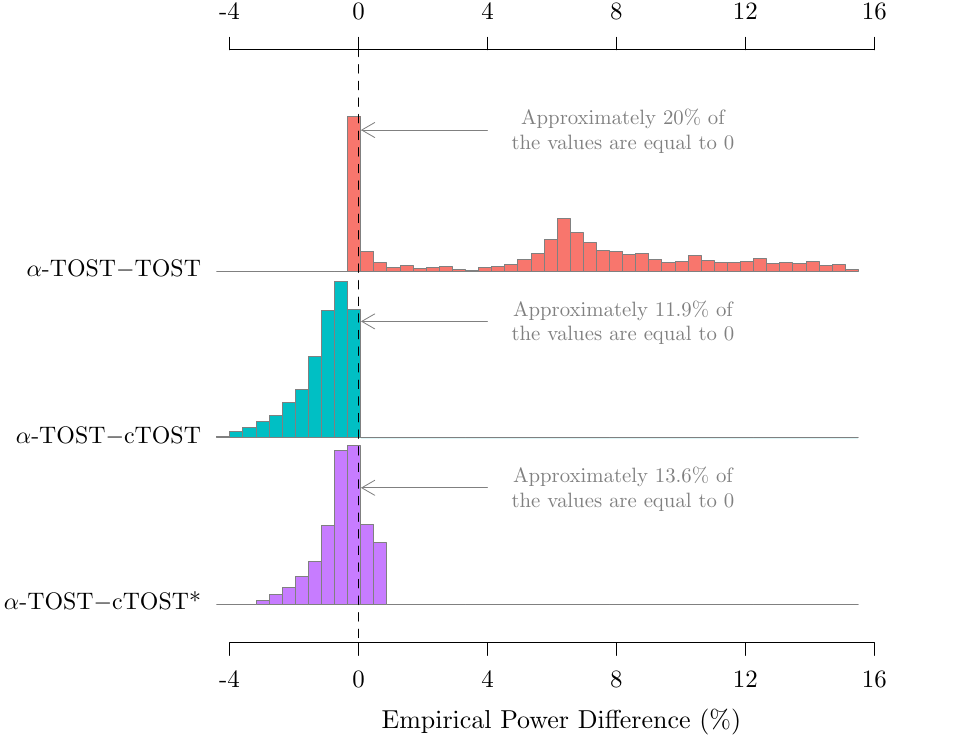}
        \caption{Histograms reporting the difference in power at $\theta=0$ for TOST, cTOST and cTOST* relative to $\alpha$-TOST when varying $\sigma_1$ and $\nu_2$.}
        \label{fig:hist_pow}
\end{figure}

This pattern is further illustrated in Figure~\ref{fig:hist_tier}, which reports the empirical test size across these simulation scenarios. 
The cTOST* procedure yields a test size that typically remains within the simulation error tolerance (displayed as a gray region around $\alpha_0=5\%$). 
On the other hand, cTOST occasionally demonstrates liberal behavior.
While TOST shows extreme conservativeness, with 33\% of the occurrences leading to a size of zero, $\alpha$-TOST 
moderately improves upon this limitation but remains substantially conservative.
Finally,  Figure~\ref{fig:hist_pow} illustrates the difference in power at $\theta=0$ relative to the $\alpha$-TOST.
These results confirm that cTOST achieves superior power performance. Although cTOST* leads to slightly reduced power compared to cTOST as a trade-off for more precise size control, it typically outperforms both TOST and $\alpha$-TOST procedures in terms of statistical power.


\clearpage

\cleardoublepage
\newpage
\section{Extended multivariate simulation results}
\label{appendix:sim_mvt_main}

\begin{figure}
        \centering
        \includegraphics[width=1\textwidth]{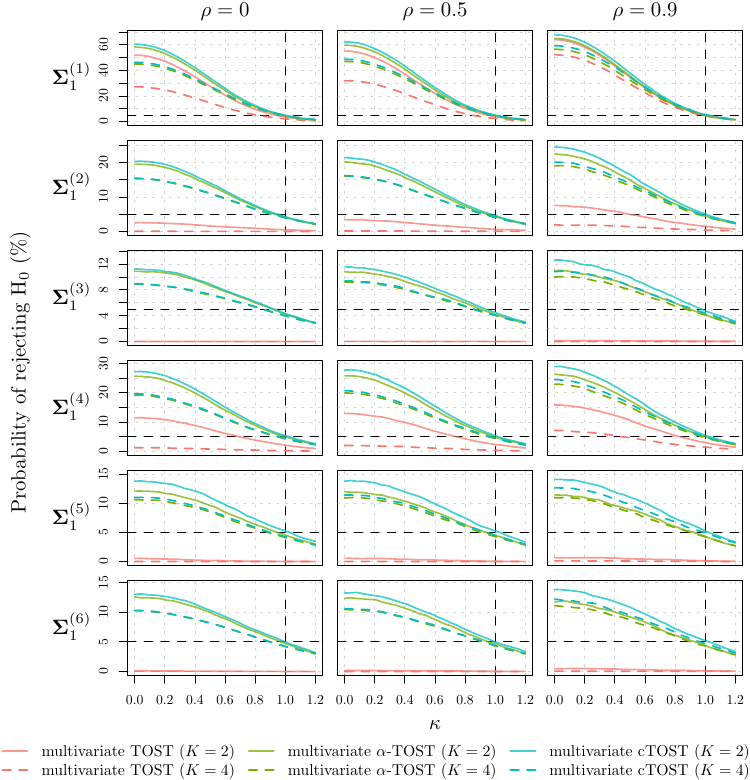}
        \caption{Probability of rejecting the null hypothesis as a function of $\kappa$, for the multivariate TOST, the multivariate $\alpha$-TOST, and the multivariate cTOST across different levels of correlation $\rho$ (columns) and covariance matrices (rows), for the bivariate (solid lines) and 4-variate (dashed lines) settings.}
  \label{fig:mvt_power_sim_ext}
\end{figure}

For the multivariate simulation settings of Section~\ref{sec:simulation_multiv}, Figure~\ref{fig:mvt_power_sim_ext} compares the performance of multivariate TOST, $\alpha$-TOST and cTOST.
Across all scenarios, the multivariate cTOST outperforms multivariate $\alpha$-TOST in terms of power, while more accurately controlling the test size at the nominal level.

\end{document}